\documentclass[aps,prb,reprint,superscriptaddress]{revtex4-1}
\usepackage{amssymb,amsmath,mathptmx}
\usepackage{graphicx}
\usepackage{dcolumn}
\usepackage{bm}
\usepackage{hyperref}
\usepackage{xcolor}
\usepackage[utf8x]{inputenc}

\begin{document}

\title{Ultralong-distance quantum correlations in three-terminal
  Josephson junctions}

\author{R\'egis M\'elin}

\affiliation{Univ. Grenoble-Alpes, CNRS, Grenoble INP\thanks{Institute
    of Engineering Univ. Grenoble Alpes}, Institut NEEL, 38000
  Grenoble, France}
\date{\today}

\begin{abstract}
  {The production of entangled pairs of electrons in
    ferromagnet-superconductor-ferromagnet or normal
    metal-superconductor-normal metal three-terminal structures has
    aroused considerable interest in the last twenty years. In these
    studies, the distance between the contacts is limited by the
    zero-energy superconducting coherence length. Here, we demonstrate
    nonlocality and quantum correlations in voltage-biased
    three-terminal Josephson junctions over the ultralong distance
    that exceeds the superconducting coherence length by orders of
    magnitude.  The effect relies on} the interplay between the
  time-periodic Floquet-Josephson dynamics, Cooper pair splitting and
  long-range coupling similar to the two-terminal Tomasch effect.  We
  find cross-over between the ``Floquet-Andreev quartets'' (if {the
    spatial separation} is smaller than the superconducting coherence
  length), and the ``ultralong-distance Floquet-Tomasch clusters of
  Cooper pairs'' if {the separation exceeds the superconducting
    coherence length, possibly reaching the same $\simeq 30\,\mu$m as
    in the Tomasch experiments}. The effect can be detected with
  DC-transport and zero-frequency quantum current-noise
  cross-correlation experiments, and it can be used for fundamental
  studies of superconducting quasiparticle quantum coherence in the
  circuits of quantum engineering.
\end{abstract}
\maketitle

\section{Introduction}

The recent developments in the field of quantum engineering allow
manipulation of long-range quantum objects with a few degrees of
freedom. Superconductivity {is a platform} for fundamental studies of
large-scale quantum systems \cite{Kouznetsov,Clarke1,Clarke2,Devoret}
and for assembling quantum processors \cite{Martinis}. Superconducting
quasiparticles can generally propagate over the entire sample and
quasiparticle poisoning
\cite{Martinis2009,deVisser2011,Lenander2011,Rajauria2012,Wenner2013,Riste2013,LevensonFalk2014,Nazarov-qp}
turns out to severely limit the range of quantum mechanical coherence
in superconductors. Superconducting devices with three or more
terminals could naturally be used for fundamental studies of coherent
quasiparticle propagation. Propagation over $R_0$ across one of the
superconducting leads, say $S_c$, trivially requires two interfaces,
one with $S_a$ and the other one with $S_b$, thus forming
$S_a$-$S_c$-$S_b$ double Josephson junction where $S_a$ and $S_b$ are
laterally connected to $S_c$ at distance $R_0$. The field of
multiterminal Josephson junctions
\cite{Freyn,Melin1,Jonckheere,FWS,Sotto,engineering,papierI,papierII,Josephson-dc,Pillet,Pillet2,Scherubl,Nazarov-PRR,Nazarov-PRB-AM,Lefloch,Heiblum,Kim,multiterminal-exp1,multiterminal-exp2,multiterminal-exp3,multiterminal-exp4,multiterminal-exp5,multiterminal-exp6,multiterminal-exp7,Levchenko1,Levchenko2}
has recently been enriched with the discovery of nontrivial topology
\cite{Nazarov1,Nazarov2,topo0,topo1,topo2,topo3,topo4,topo5,Berry} and
topology in the time-periodic Floquet dynamics
\cite{Feinberg1,Feinberg2,topo1_plus_Floquet}.

In view of these recent contributions, {we address here the
  fundamental question of the range of nonlocality and quantum
  correlations in the three-terminal devices formed with the two
  Josephson junction oscillators} $S_a$-$S_c$ and $S_c$-$S_b$ sharing
the grounded $S_c$. In spite of the well-known classical
synchronization of macroscopic Josephson junction circuits
\cite{Nerenberg1,Nerenberg2}, the present paper surprisingly
demonstrates mesoscopic quantum correlations in three-terminal
Josephson junctions at the ``ultralong-distance'' that exceeds the
{superconducting coherence} length $\xi_{ball}(0)$ by orders of
magnitude.

\begin{figure*}[htb]
  \centerline{\includegraphics[width=.9\textwidth]{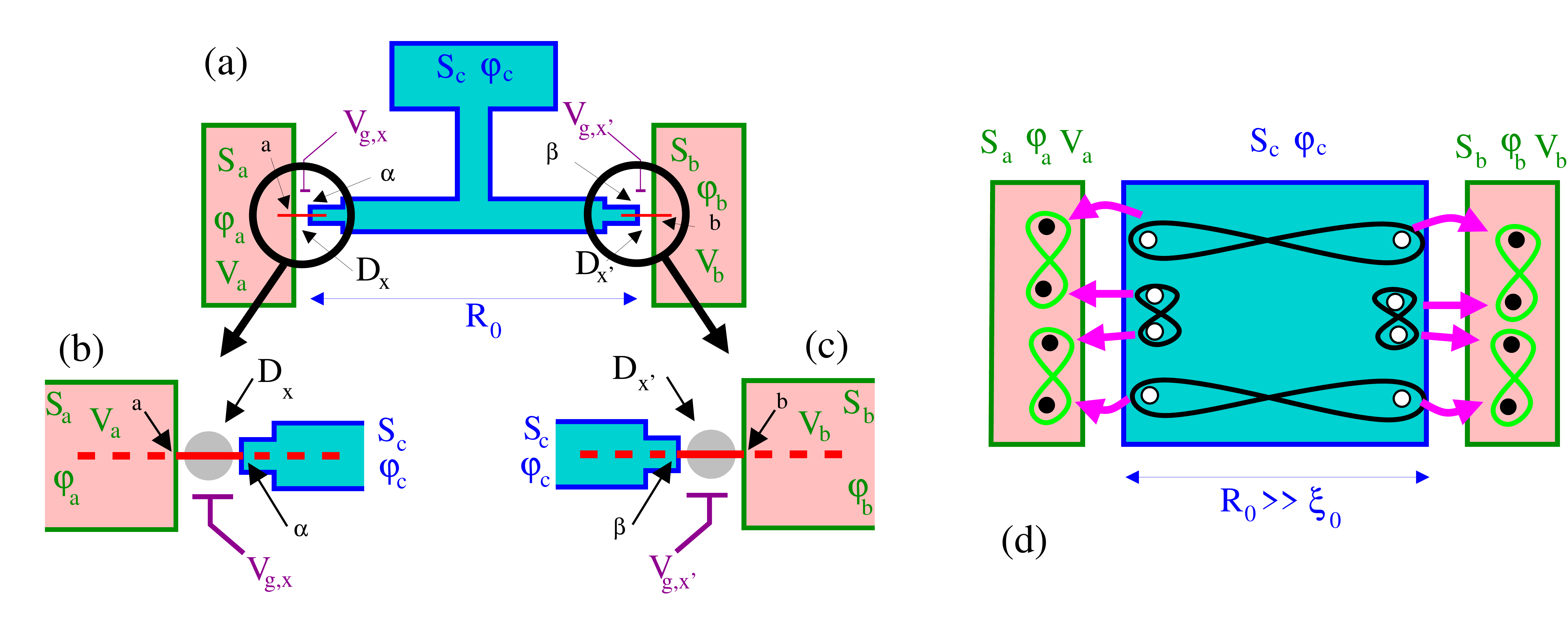}}
    \caption{{\it The device considered in the paper}, consisting of
      the three-terminal $S_a$-$D_x$-$S_c$-$D_{x'}$-$S_b$ Josephson
      junction biased on the quartet line at the voltages
      $(V_a,\,V_c,\,V_b)=(V,\,0,\,-V)$. The two quantum dots $D_x$ and
      $D_{x'}$ make distance $R_0$ between them. Panel a features the
      entire device, and panels b and c show enlargements around the
      regions of the $D_x$ and $D_{x'}$ quantum dots formed with
      semiconducting nanowires \cite{Heiblum}, with the gate voltages
      $V_{g,x}$ and $V_{g,x'}$. Panel d shows schematically the Cooper
      pair cluster of the ultralong-distance Floquet-Tomasch octets,
      see also figure~\ref{fig:octets}. The separation $R_0$ between
      the contacts can be much larger than the zero-energy coherence
      length $\xi_0 \equiv \xi_{ball}(0)$.
    \label{fig:device}
  }
\end{figure*}

Specifically, we consider a $S_a$-dot-$S_c$-dot-$S_b$ three-terminal
Josephson junction made with the BCS superconductors $S_a$, $S_b$ and
$S_c$ and two quantum dots (see
figures~\ref{fig:device}a,~\ref{fig:device}b
and~\ref{fig:device}c). This physical system has the following
features: (i) The time-periodic Floquet-Josephson dynamics with single
characteristic frequency if biasing is at commensurate voltages
\cite{Freyn,Melin1,Jonckheere,FWS,Sotto,engineering,papierI,papierII,Josephson-dc};
(ii) The nonlocal electron-hole or hole-electron conversions, {\it
  i.e.} Cooper pair splitting
\cite{exp-CPBS1,exp-CPBS2,exp-CPBS3,exp-CPBS4,exp-CPBS5,exp-CPBS6,exp-CPBS7,exp-CPBS8,theory-CPBS1,theory-CPBS2,theory-CPBS3,theory-CPBS4,theory-CPBS4-bis,theory-CPBS4-ter,theory-CPBS5,theory-CPBS6,theory-CPBS7,theory-CPBS8,theory-CPBS9,theory-CPBS10,theory-CPBS11,theory-noise8,theory-noise9,theory-noise11}
; (iii) The long-range quasiparticle propagation above the gap between
the two remote quantum dots separated by the distance $R_0$. Then, we
demonstrate that (i), (ii) and (iii) automatically imply large-scale
quantum-mechanical clusters of Cooper pairs between the constituting
$S$-dot-$S$ junctions, even if the distance $R_0$ between them is much
larger than the zero-energy {superconducting coherence}
length~$\xi_{ball}(0)$, {\it i.e.} if $R_0\gg\xi_{ball}(0)$. These
clusters can be viewed as being ``the elementary quantum particles''
that are exchanged between the two Floquet-Josephson junctions in a
three-terminal configuration. Figure~\ref{fig:device}d features
real-space representation of the lowest-order four-Cooper pair cluster
corresponding to the ``ultralong-distance Floquet-Tomasch octets''.

In the absence of bias voltage, all superconducting leads are grounded
and the three-terminal $S_a$-$S_c$-$S_b$ Josephson junction
\cite{Freyn,Pillet,Pillet2,Scherubl,Nazarov-PRR,Nazarov-PRB-AM} can be phase-biased
with appropriate superconducting loops.  The Andreev bound states
\cite{Andreev,Bretheau1,Bretheau2,Schindele,Olivares,Janvier,Gramich1,Bretheau3,Gramich2,Dassonneville,Tosi}
are then coupled by the overlapping evanescent Bogoliubov-de Gennes
wave-functions at a double interface, forming ``Andreev molecules''
with avoided crossings in their spectra, see
Refs.~\onlinecite{Pillet,Pillet2}.  At equilibrium, nonlocality is
limited by the superconducting coherence length $\xi_{ball}(0)$ as a function
of the distance $R_0$ between the $S_a$-$S_c$ and $S_c$-$S_b$
interfaces \cite{Pillet,Pillet2,Scherubl,Nazarov-PRR,Nazarov-PRB-AM}.

{We note that a DC-Josephson-like resonance appears if
  the three superconducting terminals $(S_a,\,S_c,\,S_b)$ are biased
  on the ``quartet line'' \cite{Freyn}:
\begin{equation}
  \label{eq:quartet-line}
  (V_a,V_c,V_b)=(V,\,0,\,-V)
  .
\end{equation}
The resulting Josephson relations
$\varphi_a(t)=\varphi_a+{2eVt}/{\hbar}$,
$\varphi_b(t)=\varphi_b-{2eVt}/{\hbar}$ and $\varphi_c(t)=\varphi_c$
for the superconducting phase variables at time $t$ imply the static
``quartet phase variable'' $ \varphi_q = \varphi_a(t) + \varphi_b(t) -
2 \varphi_c(t) = \varphi_a + \varphi_b - 2 \varphi_c$
\cite{Freyn}. This yields the quartet current-phase relation
$I_{c,q}\sin\varphi_q$ in the limit of low values of the contact
transparencies.  The three recent experiments of the Grenoble
\cite{Lefloch}, Weizmann Institute \cite{Heiblum} and Harvard
\cite{Kim} groups show all signs of compatibility with the theory of
the quartets
\cite{Freyn,Melin1,Jonckheere,Sotto,FWS,engineering,papierI,papierII},
in addition to other experiments
\cite{multiterminal-exp1,multiterminal-exp2,multiterminal-exp3,multiterminal-exp4,multiterminal-exp5,multiterminal-exp6,multiterminal-exp7}
on multiterminal Josephson junctions. The reason why some experiments
report the quartets while others do not is maybe a complex matter of
the materials and geometry. }

{The present paper focuses on the range of the quartets at finite bias
  voltage $V$ being a fraction of the superconducting gap $\Delta$.}
{Concerning propagation across $S_c$ between the two Josephson
  junctions, the Tomasch effect was experimentally shown in
  Refs.~\onlinecite{Tomasch1,Tomasch2,Tomasch3} to produce
  oscillations in the density of states of the superconducting
  quasiparticles in a two-terminal configuration, as a result of the
  finite superconducting film thickness reaching $33.2\,\mu$m in the
  experimental Ref.~\onlinecite{Tomasch3}. The ``Tomasch effect''
  \cite{Tomasch1,Tomasch2,Tomasch3} and the model proposed by McMillan
  and Anderson \cite{McMillan-Anderson} provide sensitivity on the
  thin film boundary conditions, corresponding to the two-terminal
  nonlocal density-phase response, see also the contribution of
  Wolfram and Lehman \cite{Wolfram}. The here considered
  three-terminal ``Floquet-Tomasch effect for the Cooper pair
  clusters'' couples one junction to the phase drop at the other
  junction according to the nonlocal current-phase response and it
  does not involve the same microscopic quantum process as the
  three-terminal density of state oscillations. The former is
  DC-current current response and the latter corresponds to AC-density
  oscillations. Nonlocality and quantum correlations are obtained in
  the Floquet-Tomasch effect over the ultralong-distance
  $R_0\gg\xi_{ball}(0)$ that is orders of magnitude larger than at
  equilibrium.}

{This ultralong-distance effect contrasts with the $F_aSF_b$
  ferromagnet-superconductor-ferromagnet and the $N_aSN_b$ normal
  metal-superconductor-normal metal beam splitters, where nonlocality
  and quantum correlations are limited by the superconducting
  coherence length $\xi_{ball}(0)$, see for instance
  Refs.~\onlinecite{exp-CPBS1,exp-CPBS2,exp-CPBS3,exp-CPBS4,exp-CPBS5,exp-CPBS6,exp-CPBS7,exp-CPBS8,theory-CPBS1,theory-CPBS2,theory-CPBS3,theory-CPBS4,theory-CPBS4-bis,theory-CPBS4-ter,theory-CPBS5,theory-CPBS6,theory-CPBS7,theory-CPBS8,theory-CPBS9,theory-CPBS10,theory-CPBS11,theory-noise8,theory-noise9,theory-noise11}.}

\begin{figure*}[htb]
  \begin{minipage}{.7\textwidth}
    \includegraphics[width=.8\textwidth]{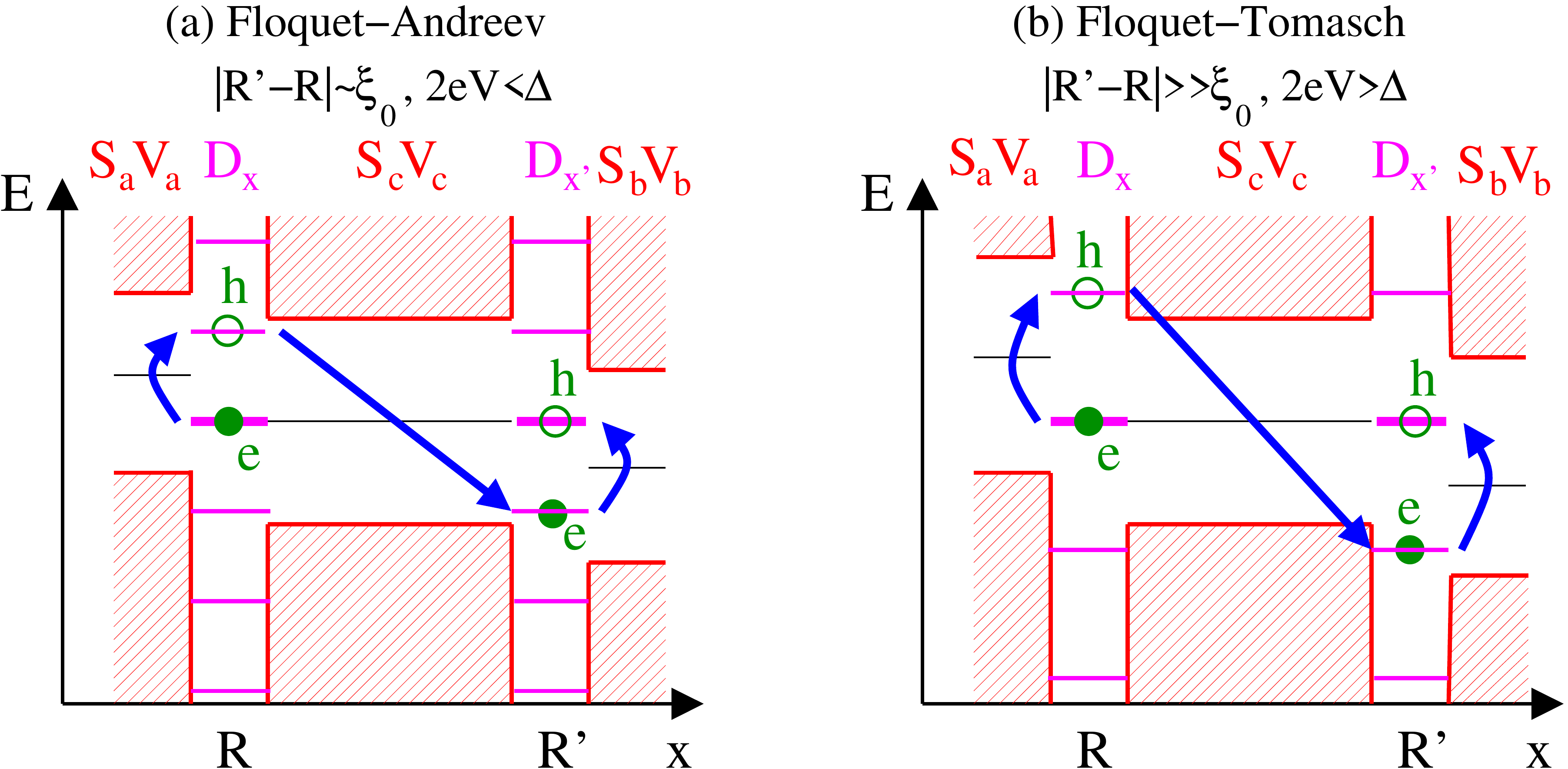}
  \end{minipage}
  \begin{minipage}{.28\textwidth}
    \caption{Schematic Floquet-Andreev pair amplitude for $2eV<\Delta$
      (a) and Floquet-Tomasch pair amplitude for $2eV>\Delta$
      (b). Both panels show conversion of a spin-up electron (e) on
      dot $D_x$ as a hole in the spin-down band (h) on dot
      $D_{x'}$. Floquet-Andreev on panel a involves subgap propagation
      over $R_0=|{\bf R}'-{\bf R}|\alt \xi_{ball}(0)$ and Floquet-Tomasch on
      panel b implies propagation above the gap if $R_0\gg \xi_{ball}(0)$.
      \label{fig:diagrams-pair-amplitude}
    }
  \end{minipage}
\end{figure*}

\begin{figure*}[htb]
  \begin{minipage}{.7\textwidth}
    \includegraphics[width=.8\textwidth]{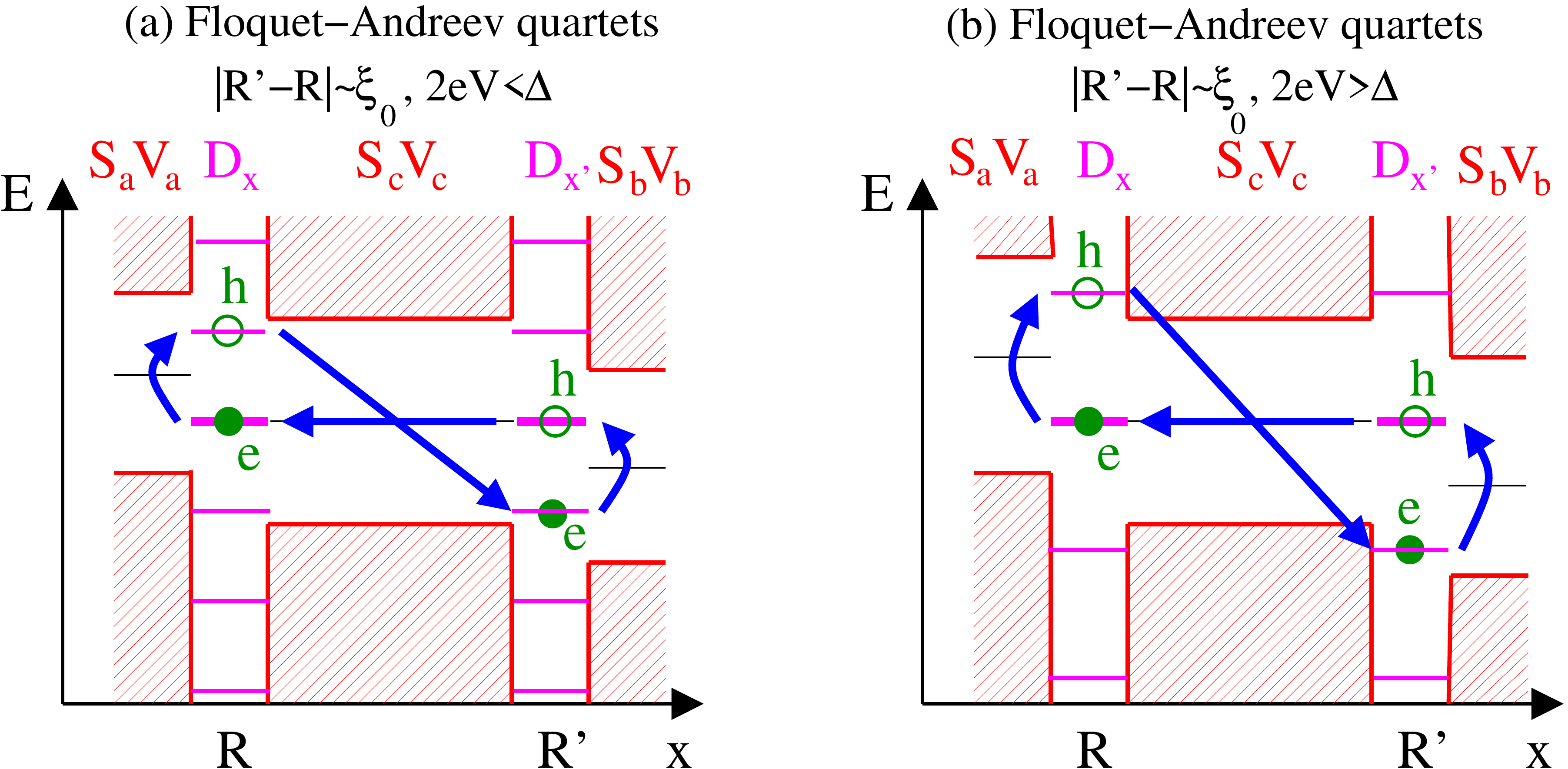}
  \end{minipage}
  \begin{minipage}{.28\textwidth}
    \caption{Schematic Floquet-Andreev DC-transport diagrams for
      $2eV<\Delta$ (a) and $2eV>\Delta$ (b). Both processes correspond to
      ``closing'' the pair amplitude diagrams on
      figure~\ref{fig:diagrams-pair-amplitude} by addition of propagation
      at zero energy from $D_{x'}$ to $D_x$, thus forming a process
      contributing to the DC-current. Both panels have distance $R_0=|{\bf
        R}'-{\bf R}|\alt \xi_{ball}(0)$ between $D_x$ and $D_{x'}$, and
      $\varphi_q$-quartet phase sensitivity.
      \label{fig:diagrams-current-and-noise}
    }
  \end{minipage}
\end{figure*}

\begin{figure}[htb]
  \includegraphics[width=.6\columnwidth]{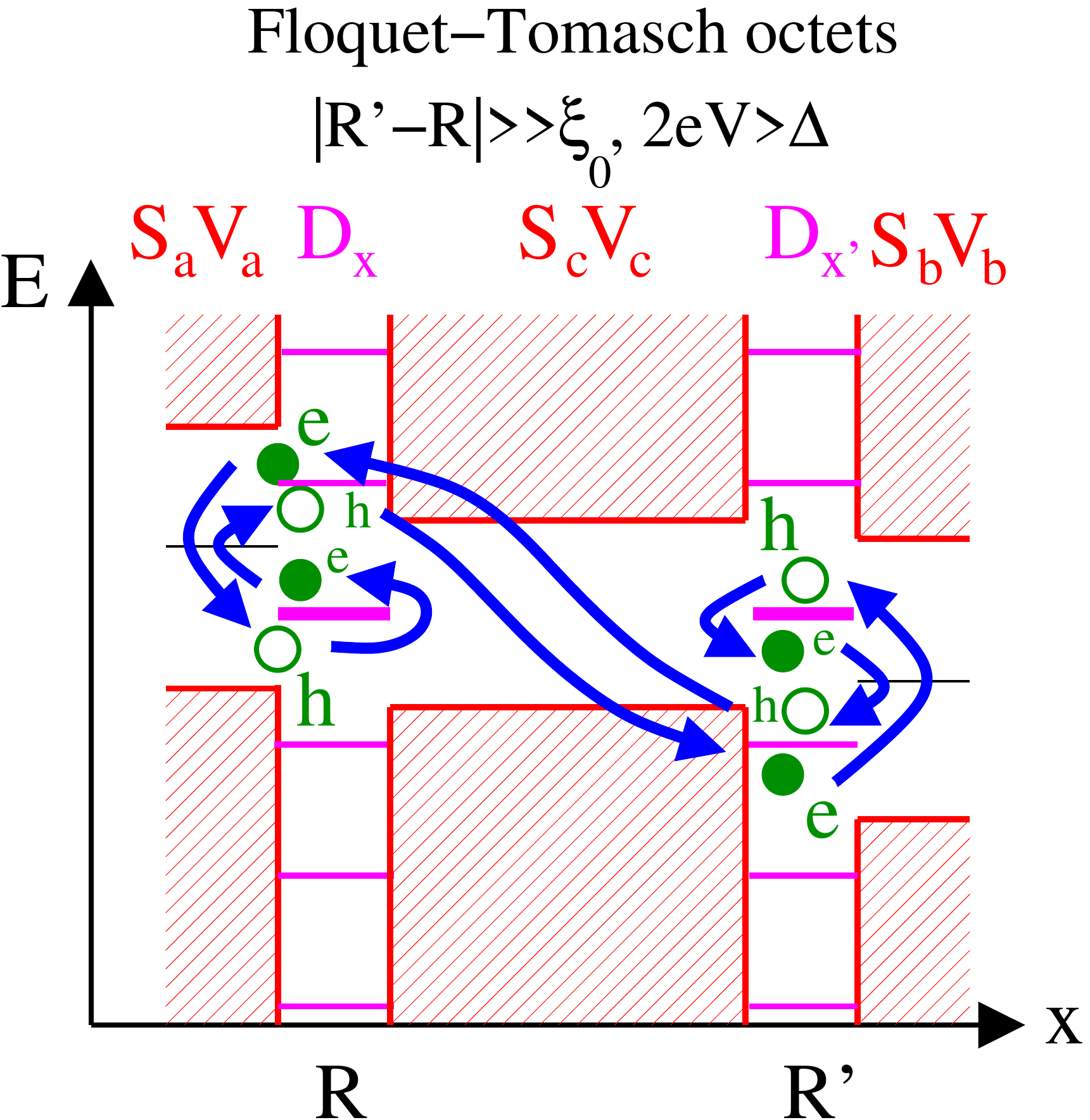}
  \caption{{\it Schematic higher-order ultralong-distance
      Floquet-Tomasch octet DC-current diagram for $2eV>\Delta$.} This
    process takes two pairs from $S_a$ at $V_a=+V$, two pairs from
    $S_b$ at $V_b=-V$, split two of them and locally transfers the two
    others into the grounded $S_c$, thus with $2\varphi_q$-quartet
    phase sensitivity. The distance between $D_x$ and $D_{x'}$ is
    $R_0=|{\bf R}'-{\bf R}|\gg \xi_{ball}(0)$. Comparing to
    figure~\ref{fig:diagrams-current-and-noise}, we deduce cross-over
    from the $\varphi_q$-quartets to the $2\varphi_q$-octets as $R_0$
    is increased from $R_0\alt\xi_{ball}(0)$ (see
    figure~\ref{fig:diagrams-current-and-noise}b) to $R_0\gg\xi_{ball}(0)$ (on
    the present figure). This process is also shown schematically on
    figure~\ref{fig:device}d.
    \label{fig:octets}
  }
\end{figure}

The paper is organized as follows. The physical picture is presented
in section~\ref{sec:physical-picture}. The model and methods are
presented in section~\ref{sec:model-and-methods}.  Analytical model
calculations are presented in section~\ref{sec:reduction-0D}.
Section~\ref{sec:numerical-results} deals with presentation of the
numerical results. Perspectives on noise measurements are discussed in
section~\ref{sec:perspectives}. Summary of the paper is provided in
section~\ref{sec:conclusions}.

\section{Physical picture}
\label{sec:physical-picture}

{ We first present the basics of Cooper pair splitting
  and nonlocality limited by the superconducting coherence length, see
  Refs.~\onlinecite{exp-CPBS1,exp-CPBS2,exp-CPBS3,exp-CPBS4,exp-CPBS5,exp-CPBS6,exp-CPBS7,exp-CPBS8,theory-CPBS1,theory-CPBS2,theory-CPBS3,theory-CPBS4,theory-CPBS4-bis,theory-CPBS4-ter,theory-CPBS5,theory-CPBS6,theory-CPBS7,theory-CPBS8,theory-CPBS9,theory-CPBS10,theory-CPBS11,theory-noise8,theory-noise9,theory-noise11}. The
  range of Cooper pair splitting is introduced in
  section~\ref{sec:range-CPS} for three-terminal $F_aSF_b$ and
  $N_aSN_b$ devices.  Next, we proceed further in section~\ref{sec:FT}
  with the ultralong-distance Floquet-Tomasch effect in a
  three-terminal $S_a$-$D_x$-$S_c$-$D_{x'}$-$S_b$ Josephson junction,
  where $D_x$ and $D_{x'}$ denote the two quantum dots.}

\subsection{Nonlocality of Cooper pair splitting}
\label{sec:range-CPS}

This subsection introduces nonlocality and quantum correlations in a
three-terminal {$F_aSF_b$ or} $N_aSN_b$ device, in
connection with Cooper pair splitting, see
Refs.~\onlinecite{exp-CPBS1,exp-CPBS2,exp-CPBS3,exp-CPBS4,exp-CPBS5,exp-CPBS6,exp-CPBS7,exp-CPBS8,theory-CPBS1,theory-CPBS2,theory-CPBS3,theory-CPBS4,theory-CPBS4-bis,theory-CPBS4-ter,theory-CPBS5,theory-CPBS6,theory-CPBS7,theory-CPBS8,theory-CPBS9,theory-CPBS10,theory-CPBS11,theory-noise8,theory-noise9,theory-noise11}.

Andreev reflection \cite{Andreev} at a normal metal-superconductor
($NS$) interface converts the supercurrent carried by Cooper pairs in
$S$ into normal current in $N$. Namely, spin-up electron from $N$ is
Andreev-reflected {as a spin-down hole and a Cooper pair} is
transmitted into the condensate. The semiclassical trajectories of
the incoming electron and outgoing hole are separated on the $NS$
interface by less than the {superconducting coherence} length
$\xi_{ball}(0)$, which is why Andreev reflection is nonlocal at the
scale of the superconducting coherence length.

The experimental evidence
\cite{exp-CPBS1,exp-CPBS2,exp-CPBS3,exp-CPBS4,exp-CPBS5,exp-CPBS6,exp-CPBS7,exp-CPBS8}
for the theoretical prediction of nonlocal Andreev reflection
\cite{theory-CPBS1,theory-CPBS2,theory-CPBS3,theory-CPBS4,theory-CPBS4-bis,theory-CPBS4-ter,theory-CPBS5,theory-CPBS6,theory-CPBS7,theory-CPBS8,theory-CPBS9,theory-CPBS10,theory-CPBS11,theory-noise8,theory-noise9,theory-noise11}
involves three-terminal configurations, such as the above mentioned
$F_aSF_b$ or $N_aSN_b$ devices.

Regarding the range of Cooper pair splitting in three-terminal
$F_aSF_b$ and $N_aSN_b$ devices, the zero-energy {superconducting coherence} length
$\xi_{ball}(0)$ is given by
\begin{equation}
  \label{eq:xi-ball-0}
  \xi_{ball}(0)=\frac{\hbar v_F}{\Delta}
\end{equation}
in the ballistic limit, where $v_F$ is the Fermi velocity. This ``size
of a Cooper pair'' is energy/frequency-$\omega$-sensitive:
\begin{equation}
  \label{eq:xi-ball-omega}
  \xi_{ball}(\omega-i\eta_S)=\frac{\hbar
    v_F}{\sqrt{\Delta^2-(\omega-i\eta_S)^2}} ,
\end{equation}
where $v_F$ is the Fermi velocity. Eq.~(\ref{eq:xi-ball-omega})
diverges as the energy $\omega$ goes to the superconducting gap
$\Delta$, see also Ref.~\onlinecite{Madrid} for the nonlocal
conductance ${\cal G}_{a,b}=\partial I_a/\partial V_b$ at arbitrary
bias voltage $V_b$ with respect to the superconducting gap.

\subsection{Ultralong-distance Floquet-Tomasch effect}
\label{sec:FT}

The introduction of ``the Feynman diagrams'' in calculations of the
light-matter interaction was not only useful to represent the quantum
processes, but it also yielded considerable shortcuts in the
calculation of those scattering amplitudes. Here, the diagrams yield
intuitive explanations and simple physical pictures for the numerical
results presented in section~\ref{sec:numerical-results}. Those
diagrams represent the time-evolution of the electrons, holes and the
conversions between them, scattering back and forth between the
different interfaces.

This subsection considers nonlocality in the
$S_a$-$D_x$-$S_c$-$D_{x'}$-$S_b$ three-terminal Josephson junction on
figures~\ref{fig:device}a, ~\ref{fig:device}b and ~\ref{fig:device}c
which is biased according to Eq.~(\ref{eq:quartet-line}) in a
{voltage-$V$ range} that is significant fraction of the
superconducting gap {$\Delta$, typically}
$eV\sim\Delta/2$. Specifically, we detail the microscopic processes,
starting with the nonlocal pair amplitude, and next proceeding further
with the Floquet-Andreev and the Floquet-Tomasch contributions to the
current, finally uncovering the ultralong-distance Floquet-Tomasch
octets. We demonstrate in Appendix~\ref{app:connection} that the
Floquet-Tomasch effect for the current of pairs in a three-terminal
Josephson junction and the two-terminal density of state oscillations
in the Tomasch effect
\cite{Tomasch1,Tomasch2,Tomasch3,McMillan-Anderson,Wolfram} share the
ultralong-distance nonlocality, but the corresponding quantum
processes are inequivalent. Thus, the mechanism for the two-terminal
density of state oscillations in the Tomasch effect
\cite{Tomasch1,Tomasch2,Tomasch3,McMillan-Anderson,Wolfram} cannot be
advocated to be at the origin of the ultralong-distance current of
pairs in the three-terminal Josephson junction. In the first place, in
the three-terminal configuration, the quantum processes coupling the
density of states at one contact to the pairs at the other contacts
are AC at the lowest-order in the tunneling amplitudes, and thus, they
cannot be put forward as an explanation to the calculated
three-terminal DC-current of quartets and higher-order clusters of
Cooper pairs.

Figures~\ref{fig:diagrams-pair-amplitude}a
and~\ref{fig:diagrams-pair-amplitude}b show the energy diagram for the
lowest-order pair amplitude between the quantum dots $D_x$ and
$D_{x'}$, corresponding to conversion of ``spin-up electron on the dot
$D_x$'' into ``spin-down hole on the dot $D_{x'}$''.

The processes on figures~\ref{fig:diagrams-pair-amplitude}a
and~\ref{fig:diagrams-pair-amplitude}b start with electron-hole
conversion at the $S_a$-$D_x$-$S_c$ Josephson junction: local
Floquet-Andreev reflection {first increases the} energy
by $2eV$ ({\it i.e.}  the energy of a Cooper pair taken from the lead
$S_a$ biased at the voltage $V$). The process continues with nonlocal
propagation from $D_x$ to $D_{x'}$ across $S_c$ in the hole-electron
channel. Next, ``local'' {inverse-Floquet hole-electron
  conversion} takes place at the $S_c$-$D_{x'}$-$S_b$ Josephson
junction. In the final state, spin-down hole is produced at zero
energy on the quantum dot $D_{x'}$.

The condition $|2eV|<\Delta$ on the bias voltage $V$ (see
figure~\ref{fig:diagrams-pair-amplitude}a) implies conversion in the
hole-electron channel over the {superconducting
  coherence length~$\xi_{ball}(\omega)$, see Eq.~(\ref{eq:xi-ball-omega}).}
This subgap process is referred to as ``the Floquet-Andreev quartet
pair amplitude''.

Conversely, $|2eV|>\Delta$ on
figure~\ref{fig:diagrams-pair-amplitude}b implies nonlocal
hole-electron conversion above the gap of $S_c$. This process is
limited by the mesoscopic phase coherence length $l_\varphi$ of the
superconducting quasiparticles, and it is referred to as ``the ultralong-distance
Floquet-Tomasch pair amplitude'' [see the forthcoming
  Eqs.~(\ref{eq:l-phi-1})-(\ref{eq:l-phi-max-2})], in analogy with the
Tomasch effect \cite{Tomasch1,Tomasch2,Tomasch3,McMillan-Anderson}
mentioned above in the Introduction.

Emergence of the ultralong-distance Floquet-Tomasch pair amplitude if
$|2eV|>\Delta$ implies ultralong-distance nonlocality over $R_0 \sim
l_\varphi$, and quantum correlations in the $\varphi_q$-sensitive
current, which is considered now.

Now, we ``close the loop'' on
figures~\ref{fig:diagrams-current-and-noise}a
and~\ref{fig:diagrams-current-and-noise}b with final zero-energy
hole-electron conversion from $D_{x'}$ to $D_x$. The resulting
$\varphi_q$-sensitive Floquet-Andreev quartet current is limited by
the {superconducting coherence} length~$\xi_{ball}(0)$, independently
on whether $|2eV|<\Delta$ or $|2eV|>\Delta$.

Finally, we consider the higher-order process of the
ultralong-distance Floquet-Tomasch octets having
$\sin(2\varphi_q)$-sensitivity and range limited by $l_\varphi$.
Figure~\ref{fig:octets} shows the corresponding diagram, see also
figure~\ref{fig:device}d. Two nonlocal and two local hole-electron
conversions are involved: (i) Nonlocally from $D_x$ to $D_{x'}$ and
from $D_{x'}$ to $D_x$ across $S_c$, and (ii) Locally between each
$S_a$ and $S_b$, the $D_x$ and $D_{x'}$ quantum dots and
$S_c$. Overall, the resulting $\sin(2\varphi_q)$-sensitive octet
current appears if the distance $R_0$ between the remote
$S_a$-$D_x$-$S_b$ and $S_c$-$D_{x'}$-$S_b$ junctions reaches $R_0 \sim
l_\varphi$, such that $l_\varphi\gg\xi_{ball}(0)$. We conclude that
figure~\ref{fig:octets} provides microscopic picture for the proposed
ultralong-distance Floquet-Tomasch octets as an eight-fermion cluster
originating from four Cooper pairs, see also figure~\ref{fig:device}d.

This physical picture suggests cross-over as $R_0$ increases from
below to above $\xi_{ball}(0)$, {\it i.e.} from ``the dominant $\sin
\varphi_q$ of the Floquet-Andreev quartets over $\xi_{ball}(0)$'' to
``the dominant $\sin(2\varphi_q)$ of the ultralong-distance
Floquet-Tomasch octets over $l_\varphi$''. {A cross-over to the
  higher-order-$n$ clusters of Cooper pairs is expected as the voltage
  values is reduced below $\Delta/2n$ (with $n$ an integer).}

{We proceed further with the models and methods in
  section~\ref{sec:model-and-methods}, next with the analytical
  results in section~\ref{sec:reduction-0D} and finally the theory is
  put to the test of the numerical calculations in
  section~\ref{sec:numerical-results}.}

\begin{figure*}[htb]
  \begin{minipage}{.7\textwidth}
    \centerline{\includegraphics[width=.7\textwidth]{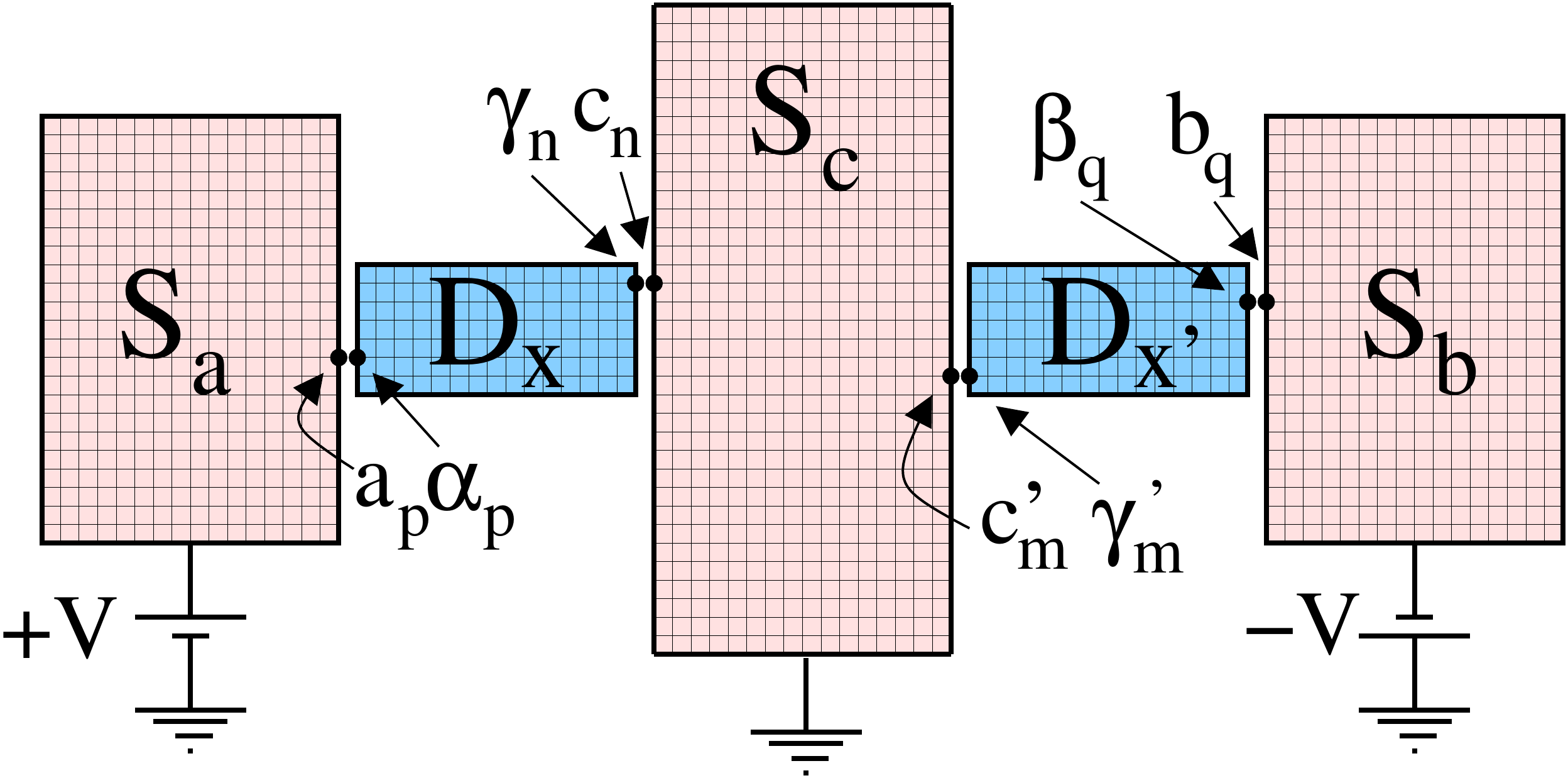}}
  \end{minipage}
  \begin{minipage}{.22\textwidth}
    \caption{Schematic tight-binding model of the considered
      $S_a$-$D_x$-$S_c$-$D_{x'}$-$S_b$ three-terminal Josephson
      junction containing two quantum dots $D_x$ and $D_{x'}$. The
      quantum dots have finite dimension on this figure. This is ``the
      model~I''.
      \label{fig:tight-binding}
    }
  \end{minipage}
\end{figure*}

\begin{figure*}[htb]
  \begin{minipage}{.7\textwidth}
    \centerline{\includegraphics[width=.7\textwidth]{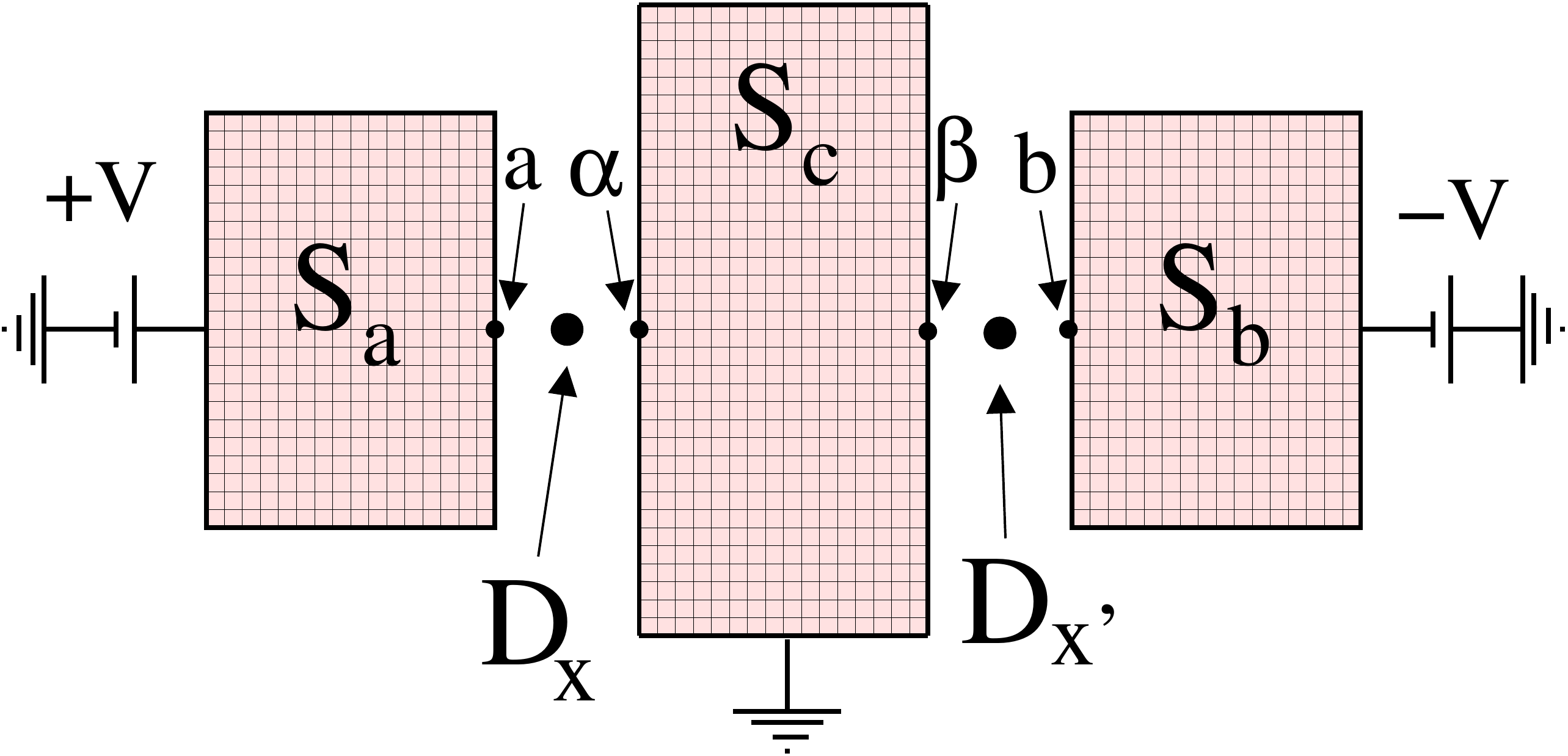}}
  \end{minipage}
  \begin{minipage}{.22\textwidth}
    \caption{Schematic tight-binding model of the considered
      $S_a$-$D_x$-$S_c$-$D_{x'}$-$S_b$ three-terminal Josephson
      junction containing two quantum dots $D_x$ and $D_{x'}$. The
      quantum dots are zero-dimensional (0D) on this figure. This is
      ``the reduced model~II''.
      \label{fig:tight-binding2}
    }
  \end{minipage}
\end{figure*}

\section{Model and methods}
\label{sec:model-and-methods}

We start in subsection~\ref{sec:geom} with a brief description of the
models used in the paper, {\it i.e.}  the geometry and the
Hamiltonians. Next, we present in subsection~\ref{sec:connection} a
central ingredient of the model, {\it i.e.} the connection between the
Dynes parameter and the mesoscopic phase coherence length of the
superconducting quasiparticles. The methods are mentioned in
section~\ref{sec:themethods}.

\subsection{Geometry and Hamiltonians}
\label{sec:geom}

Now, we present the geometry and the Hamiltonians.

Figures~\ref{fig:device}a, \ref{fig:device}b and \ref{fig:device}c
show the device geometry: the T-shaped grounded superconducting lead
$S_c$ connected {\it via} the two quantum dots $D_x$ and $D_{x'}$ to
$S_a$ and $S_b$ biased at $V_{a,b}=\pm V$. Those figures represent
quasi-one-dimensional semiconducting nanowire quantum dots similar to
Ref.~\onlinecite{Heiblum}. The distance between $D_x$ and $D_{x'}$ is
denoted by $R_0=|{\bf R}'-{\bf R}|$.

Now, we provide the Hamiltonians. The BCS Hamiltonian of each infinite
superconducting lead with gap $\Delta$ and phase $\varphi$ is given by
\begin{eqnarray}
  \label{eq:H-BCS1}
        {\cal H}_{BCS}&=&-W \sum_{\langle i,j \rangle} \sum_\sigma
        \left(c_{i,\,\sigma}^+ c_{j,\,\sigma}+ c_{j,\,\sigma}^+
        c_{i,\,\sigma}\right) \\&-& \Delta \sum_i \left(e^{i\varphi}
        c_{i,\,\uparrow}^+ c_{i,\,\downarrow}^+ + e^{-i\varphi}
        c_{i,\,\downarrow} c_{i,\,\uparrow}\right) ,
        \label{eq:H-BCS2}
\end{eqnarray}
where, again, $\sigma=\uparrow,\,\downarrow$ is the projection of the
spin along the quantization axis, and $\varphi$ takes the values
$\varphi_a$, $\varphi_b$ or $\varphi_c$ according to which of the
superconducting lead $S_a$, $S_b$ or $S_c$ is considered. The notation
$\langle i,j \rangle$ in Eq.~(\ref{eq:H-BCS1}) stands for pairs of
neighboring sites on a three-dimensional (3D) tight-binding lattice,
and the label $i$ in Eq.~(\ref{eq:H-BCS2}) runs over all tight-binding
sites.

The tunnel Hamiltonian couples the tight-binding sites on both sides of the
contacts:
\begin{equation}
  \label{eq:HT}
  {\cal H}_T=-J \sum_{\langle i,j\rangle} \sum_\sigma
  \left(c_{i,\sigma}^+ c_{j,\sigma} +
  c_{j,\sigma}^+ c_{j,\sigma} \right)
  ,
\end{equation}
where $\langle i,j\rangle$ in Eq.~(\ref{eq:HT}) denotes the pairs of
corresponding tight-binding sites on both sides of the two-dimensional
(2D) interface.

{The Hamiltonian of a direct-gap semiconductor making the quantum dots
  on figure~\ref{fig:tight-binding} is inspired by
  Ref.~\onlinecite{Gavoret}. We take the following Hamiltonian in the
  infinite 3D bulk limit:
\begin{equation}
  \label{eq:H1}
  {\cal H}_1=\sum_{\bf q} \sum_\sigma \frac{|{\bf q}|^2}{2m_e} a_{{\bf
      q},\sigma}^+ a_{{\bf q},\sigma} - \sum_{\bf q} \sum_\sigma \left(
  E_g + \frac{|{\bf q}|^2}{2m_h}\right) b_{{\bf q},\sigma}^+ b_{{\bf
      q},\sigma} ,
\end{equation}
where $a_{{\bf q},\sigma}^+$ or $b_{{\bf q},\sigma}^+$ create
spin-$\sigma$ fermions with the wave-vector ${\bf q}$ in the
conduction or valence band, and $E_g$ is the value of the direct
gap. We will use in section~\ref{sec:reduction-0D} the fact that the
dispersion relations appearing in Eq.~(\ref{eq:H1}) have extrema
at the wave-vector $q^*=0$.

Considering the $\langle a,\alpha\rangle$ pair of tight-binding sites
making the contact at the interface between the superconductor $S_a$
and the quantum dot $D_x$ (see figure~\ref{fig:tight-binding}), the
local creation operator $c_{\alpha,\sigma}^+$ on the surface $D_x$ is
defined as a sum over the quantum numbers $n_1$ and $n_2$ of the
$a_{n_1,\sigma}^+$ and $b_{n_2,\sigma}^+$ creation operators
associated to both conduction and valence band respectively:
\begin{equation}
  \label{eq:c+}
  c_{\alpha,\sigma}^+
  =\sum_{n_1} 
\varphi_{a,\,n_1}^*({\bf R}_\alpha) a_{n_1,\sigma}^+
  + \sum_{n_2} 
\varphi_{b,\,n_2}^*({\bf R}_\alpha) b_{n_2,\sigma}^+
,
\end{equation}
where we assumed a quantum dot with finite dimension, and the
tight-binding site labeled by $\alpha$ is at the space coordinate
${\bf R}_\alpha$. In Eq.~(\ref{eq:c+}), the quantum numbers $n_1$ and
$n_2$ label the states of the quantum dot with finite dimension,
possibly with irregularities in its shape, and having
Eq.~(\ref{eq:H1}) as its bulk Hamiltonian. The notations
$\varphi_{a,n_1}({\bf R}_\alpha)$ and $\varphi_{b,n_2}({\bf
  R}_\alpha)$ stand for the corresponding conduction and valence band
wave-functions.}

{The zero-dimensional (0D) quantum dot on
  figure~\ref{fig:tight-binding2} has level at zero energy. Thus, the
  corresponding Hamiltonian is ${\cal H}'_1=0$.}

{The quantum dots are connected with highly transparent
  interfaces to the leads, which is why the Coulomb interaction is
  included neither in Eq.~(\ref{eq:H1}) nor in ${\cal H}'_1=0$. For
  instance, the recent experiments \cite{multiterminal-exp7} on
  Andreev molecules
  \cite{Pillet,Pillet2,Scherubl,Nazarov-PRR,Nazarov-PRB-AM} do not
  seem to require Coulomb interactions as a central ingredient,
  because of the highly transparent interfaces.}

{Zero temperature is assumed throughout the
  paper. Nontrivial quasiparticle populations can be produced at zero
  temperature by driving normal current between two attached normal
  leads. An interesting theoretical and experimental question is to
  address whether driving normal current can result in change of sign
  of the quartet critical current, similarly to two terminals, see
  Refs.~\onlinecite{vanWees-PRB,vanWees}.  }

{The scattering approach or the Keldysh Green's functions
  \cite{Caroli} were complementary used in the past to address
  superconducting junctions, see for instance
  Refs.~\onlinecite{Averin,Cuevas,Cuevas-noise} for a single
  superconducting weak link. Both approaches have their own
  advantages. For instance, the scattering matrix calculations and the
  wave-function approach allow for semiclassical calculations, see
  Refs.~\onlinecite{Berry,Bratus}. Microscopic Green's functions
  produce efficient algorithms to address the general conditions of
  high transparencies and large current bias, see for instance
  Ref.~\onlinecite{Madrid}. In the following, we rely on the Keldysh
  Green's functions, on the basis of the algorithms that were
  developed over the last few years
  \cite{FWS,Sotto,engineering,papierII,Berry}.}

{We also implement the simplifying assumption of a ballistic
  superconductor, similarly to the McMillan-Anderson and the Wolfram-Lehman
  papers \cite{McMillan-Anderson,Wolfram} on the Tomasch effect
  \cite{Tomasch1,Tomasch2,Tomasch3}. Taking the ballistic limit yields
  considerable simplifications in the calculations, see
  below. Disorder in the superconductors could be introduced in the
  future on the basis of the Usadel equations \cite{Usadel}. Another
  possible approach is to assume perturbation theory in the strength
  of the nonlocal processes between the two quantum dots, see the
  forthcoming section~\ref{sec:reduction-0D}, and to average over
  disorder the pairs of nonlocal Green's functions connecting both
  quantum dots.  The $16$ Nambu components of the advanced-advanced
  transmission modes (see Ref.~\onlinecite{papierI}) would then have
  to be generalized to the Keldysh contour and to energy outside the
  superconducting gap.  }

\subsection{The mesoscopic phase coherence length of the superconducting quasiparticles}
  \label{sec:connection}
In this subsection, we relate the mesoscopic phase coherence length
$l_\varphi$ of the superconducting quasiparticles to the Dynes
parameter $\eta_S$\cite{FWS,Kaplan,Dynes,Pekola1,Pekola2}.

By the time-energy uncertainty relation, and by the correspondence
between the time and length scales, a characteristic length $\hbar
v_F/E_0$ is associated to any energy scale $E_0$. To the Fermi energy
$\epsilon_F$ is associated the Fermi wave-length $\lambda_F$, which is
much smaller than the superconducting coherence length $\xi_{ball}(0)$
that is related to the superconducting gap $\Delta$. The
characteristic length $l_\varphi$ is conjugate to the Dynes parameter
$\eta_S$, and it phenomenologically accounts for the
quantum-to-classical cross-over of the propagating superconducting
quasiparticles, due to inelastic scattering and energy
relaxation. Then, $l_\varphi$ is much larger than the superconducting
coherence length $\xi_{ball}(0)$, {\it i.e.}  $l_\varphi\gg
\xi_{ball}(0)$, because the Dynes parameter $\eta_S$ is much smaller
than the superconducting gap $\Delta$, {\it i.e.}  $\eta_S\ll\Delta$,
see Refs.~\onlinecite{FWS,Kaplan,Dynes,Pekola1,Pekola2}.  The length
scale $l_\varphi$ has to cross-over to its normal-state value $\hbar
v_F/\eta_S$ as the energy $\omega$ crosses-over above $\omega\agt
2\Delta$. This $l_\varphi$ naturally receives the interpretation of
defining the ``limit of the quantum world'' as far as the
superconducting quasiparticle propagation is concerned.

Now, within this phenomenological ``Dynes picture'', we provide
analytical expressions for the mesoscopic phase coherence length
$l_\varphi$ of the superconducting quasiparticles as a function of the energy
$\omega$.

The evanescent Bogoliubov-de Gennes wave-functions decay exponentially
like $\sim \exp\left(-{R}/{\xi_{ball}(\omega-i\eta_S)}\right)$ from
the interface at the subgap energy $|\omega|<\Delta$, see also the
Green's function given by Eq.~(\ref{eq:gA-Nambu}). Then, the
superconducting coherence length $\xi_{ball}(\omega-i\eta_S)$ can be
continued to energies $|\omega|>\Delta$ outside the gap, and it has
the following real and imaginary parts:
\begin{eqnarray}
 \frac{1}{\xi_{ball}(\omega-i\eta_S)}
  &=& \mbox{Re}\left(\frac{1}{\xi_{ball}(\omega-i\eta_S)}\right)\\
&&  +  i\, \mbox{Im}\left(\frac{1}{\xi_{ball}(\omega-i\eta_S)}\right)
  ,\nonumber
\end{eqnarray}
which yields damping and oscillations:
\begin{eqnarray}
  \nonumber
&& \exp\left(-\frac{R}{\xi_{ball}(\omega-i\eta_S)}\right) =
  \exp\left[-R \, \mbox{Re}\left(\frac{1}{\xi_{ball}(\omega-i\eta_S)}
    \right)\right] \\ && \times \exp\left[-i \, R \,
    \mbox{Im}\left(\frac{1}{\xi_{ball}(\omega-i\eta_S)}
    \ \right)\right] .
\end{eqnarray}
We define the inverse damping length as
  \begin{eqnarray}
    \label{eq:l-phi-1}
    \frac{1}{l_\varphi}&=&
    \mbox{Re}\left[\frac{1}{\xi_{ball}(\omega-i\eta_S)}\right]
    ,
  \end{eqnarray}
  with $|\omega|>\Delta$.

{ We note that $ \mbox{Re} \sqrt{\Delta^2 - (\omega-i\eta_S)^2} = \rho
  \cos(\theta/2)$ , where $\rho^2 \exp(i\theta)= \Delta^2-\omega^2
  +\eta_S^2 +2i\eta_S\omega$. Using $\cos \theta=2\cos^2(\theta/2)-1$
  leads to
  \begin{eqnarray}
&&    \mbox{Re}\sqrt{\Delta^2-(\omega-i\eta_S)^2}\\&=&\frac{1}{\sqrt{2}}
    \left\{\left[\left(\Delta^2-\omega^2+\eta_S^2\right)^2
      +4 \eta_S^2\omega^2\right]^{1/2}
    +\Delta^2-\omega^2+\eta_S^2\right\}^{1/2}
    .
    \nonumber
  \end{eqnarray}
  Assuming $\eta_S\ll \Delta$ and $|\omega|>\Delta$ yields
  \begin{eqnarray}&&\left\{
  \left(\Delta^2-\omega^2+\eta_S^2\right)^2
  +4 \eta_S^2\omega^2\right\}^{1/2}\\
  \nonumber
&&\simeq
\left|\Delta^2-\omega^2\right|
+\eta_S^2 \frac{\Delta^2+\omega^2}{\left|\Delta^2-\omega^2\right|}
,
  \end{eqnarray}
  and
  \begin{eqnarray}
&&    \left\{
  \left[\left(\Delta^2-\omega^2+\eta_S^2\right)^2
      +4 \eta_S^2\omega^2\right]^{1/2}
  +\Delta^2-\omega^2+\eta_S^2\right\}^{1/2}\\
  \nonumber
  &&\simeq
  \left\{\eta_S^2 \frac{\Delta^2+\omega^2+\left|\Delta^2-\omega^2\right|}
    {\left|\Delta^2-\omega^2\right|}\right\}^{1/2}
  ,
  \end{eqnarray}
  where we used $\Delta^2-\omega^2+\left|\Delta^2-\omega^2\right|=0$
  if $|\omega|>\Delta$. The following is deduced:
  \begin{eqnarray}
    \mbox{Re}\sqrt{\Delta^2-(\omega-i\eta_S)^2}
    \simeq \eta_S \frac{|\omega|}{[\omega^2-\Delta^2]^{1/2}}
      ,
  \end{eqnarray}
  and $1/l_\varphi$ is given by
  \begin{equation}
    \frac{1}{l_\varphi}\simeq
    \frac{\eta_S}{\hbar v_F}\times
    \frac{|\omega|}{[\omega^2-\Delta^2]^{1/2}}
    \label{eq:l-phi-2}
    .
  \end{equation}
  Then, $l_\varphi\approx l_\varphi^{max}$ if the energy $\omega$ takes
  the typical value $|\omega|\approx 2\Delta$:
  \begin{equation}
    \label{eq:l-phi-max}
    l_\varphi^{(max)}=\frac{\hbar v_F}{\eta_S}
    .
  \end{equation}
  This yields
  \begin{equation}
    \label{eq:l-phi-max-2}
    \frac{l_\varphi^{(max)}}{\xi_{ball}(0)}=\frac{\Delta}{\eta_S}
    ,
  \end{equation}
  where $l_\varphi^{(max)}$ is expressed in units of the zero-energy
  superconducting coherence length $\xi_{ball}(0)$, see
  Eq.~(\ref{eq:xi-ball-0}). The Dynes ratio $\eta_S/\Delta\ll 1$ is
  small in the experiments \cite{Kaplan,Dynes,Pekola1,Pekola2}, which
  implies the ultralong-distance effect corresponding to
  $l_\varphi^{(max)}/\xi_{ball}(0)\gg 1$ in
  Eq.~(\ref{eq:l-phi-max-2}).

  Thus, Eq.~(\ref{eq:l-phi-max-2}) supports the idea presented in the
  Introduction, {\it i.e.} within this Dynes picture, the mesoscopic
  phase coherence length $l_\varphi$ of the superconducting quasiparticles
  is orders of magnitude larger than the zero-energy superconducting
  coherence length $\xi_{ball}(0)$ in a typical energy window that can
  roughly be estimated as $|\omega|\approx 2 \Delta$.  This typical
  spectral window for emergence of the ultralong-distance
  Floquet-Tomasch effect reflects the coexistence of both features of
  the normal and superconducting states, {\it i.e.} long $l_\varphi$
  and sizeable nonlocal Andreev processes.}

  {Controlling the electromagnetic environment as in
    Ref.~\onlinecite{Pekola1} can reduce the value of the Dynes
    parameter $\eta_S$ by orders of magnitude, and produce large value
    of $l_\varphi$ according to
    Eqs.~(\ref{eq:l-phi-1})-(\ref{eq:l-phi-max-2}). This can also be used
    to rule out the coupling to the electromagnetic environment as the
    origin of the quartet line. In the previous Grenoble
    \cite{Lefloch} and Weizmann group experiments \cite{Heiblum}, a
    device fabricated with remote junctions did not produce the
    quartet line in spite of the same electromagnetic environment as
    in the device with close junctions.}

  \subsection{Methods}
\label{sec:themethods}
  The analytical and numerical calculations presented in
  sections~\ref{sec:reduction-0D} and~\ref{sec:numerical-results}
  respectively are based on the Keldysh Green's functions. Details
  about the methods are provided in Appendix~\ref{app:themethod}.
  
\section{Analytical results}
\label{sec:reduction-0D}

In this section, we assume that the quantum dots are fabricated with
direct-gap semiconductors {[see Eq.~(\ref{eq:H1})]}, and we map ``the
model~I'' on figure~\ref{fig:tight-binding} onto ``the reduced
model~II'' on figure~\ref{fig:tight-binding2}. We also provide
analytical results demonstrating the Floquet-Andreev quartets and the
ultralong-distance Floquet-Tomasch octets, discuss the absence of
dephasing in propagation between the two interface and explain why the
ultralong-distance effect appears both for the two-terminal density of
states in the Tomasch experiments \cite{Tomasch1,Tomasch2,Tomasch3},
and for the pair current in the here considered three-terminal
Josephson junction. However, the quantum processes are distinct from
each other and it turns out that the nonlocal coupling between the
density of states at one contact and the pairs at the other contact is
AC in the three-terminal Josephson junction.

Specifically, starting with the model~I {in}
figure~\ref{fig:tight-binding}, we assume that the Nambu Green's
function of each quantum dot $D_x$ or $D_{x'}$ fulfills the following
``generalized star-triangle relation'', {\it i.e.} we propose the
following for the quantum dot $D_x$:
\begin{eqnarray}
  \label{eq:star-triangle1}
  \hat{g}_{\alpha_{p_1},\,\alpha_{p_2}}&=&\tilde{g}_{\alpha_{p_1},x} \tilde{g}_{x,x}
  \tilde{g}_{x,\,\alpha_{p_2}}\\
   \hat{g}_{\alpha_{p_1},\,\gamma_{n_2}}&=&\tilde{g}_{\alpha_{p_1},x} \tilde{g}_{x,x}
   \tilde{g}_{x,\,\gamma_{n_2}}
   .
   \label{eq:star-triangle2}
\end{eqnarray}
The assumption of resonance at zero energy implies $\tilde{g}_{x,x}^A
\sim 1/(\omega-i\eta)$, see the discussion following
Eq.~(\ref{eq:g-rep-spec}) in Appendix~\ref{app:themethod}. We consider
that the quantum dots have minimum at the wave-vector $q^*=0$ in their
dispersion relation, see Eq.~(\ref{eq:H1}). We assume that the contact
dimension is small compared to $2\pi/q^*$ and that the size of the
quantum dots is small compared to the decay length of the evanescent
wave-functions on the dot. Then,
$\hat{g}_{\alpha_{p_1},\,\alpha_{p_2}}$ and
$\hat{g}_{\alpha_{p_1},\,\gamma_{n_2}}$ are roughly independent on
$p_1,\,\,p_2$ and $\tilde{g}$ is the matrix square root of the residue
in Eq.~(\ref{eq:g-rep-spec-approx}).

The Green's functions are matrices in Nambu and in the enlarged space
of the harmonics of the Josephson frequency. The labels $p_1,\,\,p_2$
running over the tight-binding sites at the interfaces are now made
implicit [see
  Eqs.~(\ref{eq:star-triangle1})-(\ref{eq:star-triangle2})].

The fully dressed
Green's function $\tilde{G}_{x,x}$ on the dot $D_x$ can be ``expanded
in nonlocality'' according to
\begin{eqnarray}
  \label{eq:Dyson-exp}
  \tilde{G}_{x,x}&=&\tilde{L}_{x,x}\\ &+&\tilde{L}_{x,x}
  \tilde{K}_{x,x'} \tilde{L}_{x',x'} \tilde{K}_{x',x}
  \tilde{L}_{x,x}\label{eq:Dyson-exp2}\\ &+&
\tilde{L}_{x,x}
  \tilde{K}_{x,x'} \tilde{L}_{x',x'} \tilde{K}_{x',x}
  \tilde{L}_{x,x}\tilde{K}_{x,x'} \tilde{L}_{x',x'} \tilde{K}_{x',x}
  \tilde{L}_{x,x} 
  \label{eq:Dyson-exp3}\\&+&...\label{eq:Dyson-exp4} 
  ,
\end{eqnarray}
where $\tilde{L}_{x,x}$ and $\tilde{L}_{x',x'}$ describe ``local''
dressing at the $S_a$-$D_x$-$S_c$ and $S_c$-$D_{x'}$-$S_b$ junctions,
and the matrices $\tilde{K}_{x,x'}$ and $\tilde{K}_{x',x}$ correspond
to nonlocal propagation from $x$ to $x'$ and from $x'$ to $x$
respectively, see Appendix~\ref{app:details}. An expansion similar to
Eqs.~(\ref{eq:Dyson-exp})-(\ref{eq:Dyson-exp4}) was previously
developed for the nonlocal conductance of $F_aSF_b$ or $N_aSN_b$ beam
splitters, see Ref.~\onlinecite{theory-CPBS11}.  Here, the small
parameter for nonlocality of the Floquet-Andreev quartets is
$\epsilon_0=\exp[-2R_0/\xi_{ball}(0)]$, due to transmission of
quasiparticles {\it via} evanescent states in the subgap energy
window. The small parameter for the Floquet-Tomasch octets is
$\epsilon_\varphi=\exp[-2R_0/l_\varphi(\omega)]$ instead of the
previous $\epsilon_0$, corresponding to propagation {\it via} plane
waves in a spectral window above the gap of $S_c$, and damping over
the mesoscopic phase coherence length $l_\varphi(\omega)$, see
Eqs.~(\ref{eq:l-phi-1})-(\ref{eq:l-phi-2}).

The first term in Eq.~(\ref{eq:Dyson-exp}) does not couple the two
quantum dots. The Keldysh component of the second term in
Eq.~(\ref{eq:Dyson-exp2}) is the following:
\begin{widetext}
  \begin{equation}
    \Sigma_{a,\,\alpha} G^{+,-}_{\alpha,a}\simeq
    \left( \Sigma_{a,\,\alpha} \tilde{g}_{\alpha,x}
    \tilde{L}_{x,x} \tilde{K}_{x,x'}
    \tilde{L}_{x',x'} \tilde{K}_{x',x}
    \tilde{L}_{x,x} \tilde{g}_{x,\,\alpha}
    \Sigma_{\alpha,a} g_{a,a} \right)^{+,-}
    .
  \end{equation}
Specifying the Nambu labels corresponding to anomalous propagation
between $D_x$ and $D_{x'}$ leads to
\begin{equation}
  \label{eq:Z1}
    \left(\Sigma_{a,\,\alpha} G^{+,-}_{\alpha,a}\right)_{(1,1)}\simeq
    \left( \Sigma_{a,\,\alpha,(1,1)} \tilde{g}_{\alpha,x,(1,1)}
    \tilde{L}_{x,x,(1,1)} \tilde{K}_{x,x',(1,2)}
    \tilde{L}_{x',x',(2,1)} \tilde{K}_{x',x,(1,2)}
    \tilde{L}_{x,x,(2,2)} \tilde{g}_{x,\,\alpha,(2,2)}
    \Sigma_{\alpha,a,(2,2)} g_{a,a,(2,1)} \right)^{+,-}
    ,
\end{equation}
{Within this
  approximation, the Floquet-Tomasch quartets and octets propagate a
  pair of nonlocal Green's functions between the two quantum dots,
  where Eq.~(\ref{eq:Z1}) also captures ``local'' dressing by multiple
  Andreev reflections at each $S$-dot-$S$ Josephson junction.}

The Floquet-Andreev quartets correspond to
  \begin{eqnarray}
    \label{eq:Andreev-quartets}
    \left(\Sigma_{a,\,\alpha} G^{+,-}_{\alpha,a}\right)_{(1,1)/(0,0)}&\simeq&
    \left( \Sigma_{a,\,\alpha,(1,1)/(0,1)} \tilde{g}_{\alpha,x,(1,1)/(1,1)}
    \tilde{L}_{x,x,(1,1)/(1,1)} \tilde{K}_{x,x',(1,2)/(1,1)}
    \tilde{L}_{x',x',(2,1)/(1,-1)}\right.\\&&\left. \tilde{K}_{x',x,(1,2)/(-1,-1)}
    \tilde{L}_{x,x,(2,2)/(-1,-1)} \tilde{g}_{x,\,\alpha,(2,2)/(-1,-1)}
    \Sigma_{\alpha,a,(2,2)/(-1,0)} g_{a,a,(2,1)/(0,0)} \right)^{+,-}
    \nonumber
    ,
  \end{eqnarray}
  where the ``$(\tau_1,\tau_2)/(n_1,n_2)$'' labels are used for the
  Nambu and Floquet labels respectively. {Both $\tilde{K}_{x,x'}$ and
    $\tilde{K}_{x',x}$ entering Eq.~(\ref{eq:Andreev-quartets}) are of
    order $\epsilon_0$ if $R_0\agt \xi_{ball}(0)$, due to the
    corresponding dominant contribution of the subgap energy window.
    Thus, $\left(\Sigma_{a,\,\alpha}
    G^{+,-}_{\alpha,a}\right)_{(1,1)}$ in Eq.~(\ref{eq:Z1}) is of
    order $(\epsilon_0)^2$.}
  
  The Floquet-Tomasch Keldysh Green's function is given by
  \begin{eqnarray}
    \label{eq:Floquet-octets}
    \left(\Sigma_{a,\,\alpha} G^{+,-}_{\alpha,a}\right)_{(1,1)/(0,0)}&\simeq&
    \left( \Sigma_{a,\,\alpha,(1,1)/(0,1)}
    \tilde{g}_{\alpha,x,(1,1)/(1,1)} \tilde{L}_{x,x,(1,1)/(1,3)}
    \tilde{K}_{x,x',(1,2)/(3,3)} \tilde{L}_{x',x',(2,1)/(3,1)}
    \tilde{K}_{x',x,(1,2)/(1,1)}\right.\\ && \left.
    \tilde{L}_{x,x,(2,2)/(1,-1)} \tilde{g}_{x,\,\alpha,(2,2)/(-1,-1)}
    \Sigma_{\alpha,a,(2,2)/(-1,0)} g_{a,a,(2,1)/(0,0)} \right)^{+,-}
    , \nonumber
  \end{eqnarray}
\end{widetext}
where $\tilde{K}_{x,x',(1,2)}$ and $\tilde{K}_{x',x,(1,2)}$ entering
Eq.~(\ref{eq:Floquet-octets}) are both of order $\epsilon_\varphi$ if
$R_0\agt l_\varphi$. Thus, $\left(\Sigma_{a,\,\alpha}
G^{+,-}_{\alpha,a}\right)_{(1,1)}$ in Eq.~(\ref{eq:Floquet-octets}) is
of order $(\epsilon_\varphi)^2$.

Eqs.~(\ref{eq:Andreev-quartets})-(\ref{eq:Floquet-octets}) imply that
the current $I_q$ on the quartet line is expressed as summation over
$c_1,\,c_2$ and $c'_1,\,c'_2$ at the $D_x$-$S_c$ and $S_c$-$D_{x'}$
interfaces respectively, see figure~\ref{fig:tight-binding}: $
I_q=\sum_{c_1,c_2,c'_1,c'_2} I_{c_1,c_2,c'_1,c'_2}$.  Then,
Eq.~(\ref{eq:gA-supra-general-ballistique}) in
Appendix~\ref{app:themethod} yields
\begin{equation}
  \label{eq:Iq-A}
I_q=\sum_{c_1,c_2,c'_1,c'_2} I'_{c_1,c_2,c'_1,c'_2}
\cos\left[k_F R_{c_1,c'_1}\right]
\cos\left[k_F R_{c_2,c'_2}\right]
.
\end{equation}
Gathering the Green's functions in a pair-wise manner \cite{HN1,HN2}
yields $I_q\simeq\sum_{c,c'} I_{c,c,c',c'}$ and
\begin{equation}
  \label{eq:I-quartet-av}
  I_q\simeq \frac{k_F}{2\pi} \int_{R_0}^{R_0+2\pi/k_F} I_q(R) dR ,
\end{equation}
where $I_q(R)$ is the
spectral current of the ``reduced model~II on
figure~\ref{fig:tight-binding2}'' at the distance $R$ between the
0D quantum dots.

Thus, the {use of direct-gap} semiconductor quantum dots {allows
  replacing} ``the multichannel contacts of the model~I '' by ``the 0D
quantum dots of the reduced model~II'' while averaging over $\psi_F$
in Eq.~(\ref{eq:gA-supra-general-ballistique}). We also singled-out
the Floquet-Andreev quartet and the ultralong-distance Floquet-Tomasch
octet contributions to the current, which supports the physical
picture of the preceding sections~\ref{sec:physical-picture}
and~\ref{sec:model-and-methods}, and the numerical results of the
forthcoming section~\ref{sec:numerical-results}.

We also note that biasing at $eV=\pm \Delta/2$
produces coinciding gap edge singularities of $S_a$ and $S_b$. This is
expected to result in large values for the quartet and octet critical
currents $I_{q,c} \sin\varphi_q$ and $I_{o,c} \sin(2\varphi_q)$, as
for perfectly transparent contacts.  The following scaling form of
$|I_{q,c}|$ and $|I_{o,c}|$ at the voltages $eV=\pm\Delta/2$ can be
conjectured:
\begin{eqnarray}
  \label{eq:Iqc}
  |I_{q,c}|&\approx& \frac{e}{\hbar} \exp\left(-\frac{2 R_0}{\xi_{ball}(0)}
  \right)\\
  |I_{o,c}|&\approx& \frac{e}{\hbar} \exp\left(-\frac{2 R_0}{l_\varphi}
  \right)
  \label{eq:Ioc}
  .
\end{eqnarray}
Both $|I_{q,c}|$ and $|I_{o,c}|$ are expected to be reduced if the
bias voltage is detuned from $\pm \Delta/e$. This Eq.~(\ref{eq:Ioc})
will further be considered in the next section on the numerical data.

\begin{figure*}[htb]
    \centerline{\includegraphics[width=.95\textwidth]{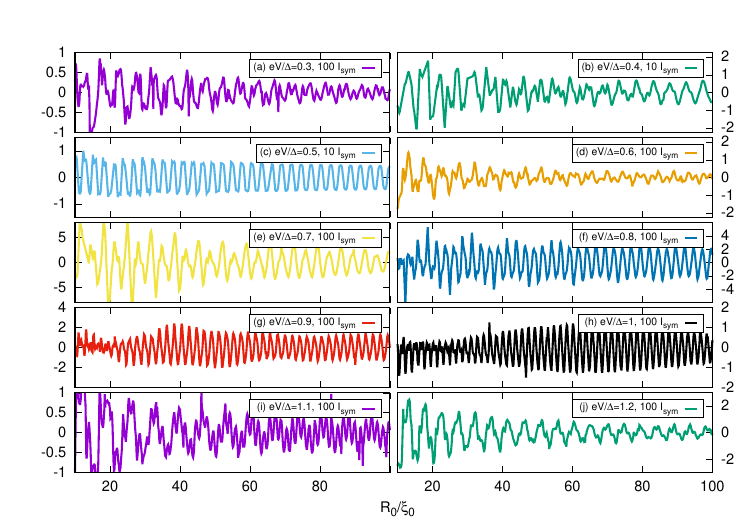}}
    \caption{{\it The ultralong-distance effect:} The figure shows the
      symmetric current $I_{sym}$ defined by Eq.~(\ref{eq:Im}) for a
      $S_a$-$D_x$-$S_c$-$D_{x'}$-$S_b$ three-terminal Josephson
      junction biased at the voltages
      $(V_a,\,V_b,\,V_c)=(V,\,-V,\,0)$. Panels a-j correspond to
      $eV/\Delta=0.3,\,0.4,\,0.5,\,0.6,\,0.7,\,0.8,\,0.9,\,1,\,1.1,\,1.2$. The
      couplings between the quantum dots and the superconducting leads
      are the following: $\Gamma_{x,a}/\Delta = \Gamma_{x',b}/\Delta =
      0.25$ and $\Gamma_{x,\alpha}/\Delta = \Gamma_{x',\beta}/\Delta =
      1$. The quartet phase is $\varphi_q/2\pi=0.1$. The Dynes
      parameter is $\eta_S / \Delta=10^{-3}${, thus with
        $l_\varphi^{(max)}=10^3 \xi_0$, where $\xi_0$ is a short
        notation for $\xi_{ball}(0)$.}  The quartet phase variable is
      set to $\varphi_q/2\pi=0.1$, see the forthcoming
      figure~\ref{fig:vs-V-1} for the $\varphi_q$-sensitivity of the
      quartet current at fixed separation
      $R_0/\xi_{ball}(0)$. The current is in units of
      $e\Delta/\hbar$.
    \label{fig:vs-R}
  }
\end{figure*}

\begin{figure*}[htb]
    \centerline{\includegraphics[width=.95\textwidth]{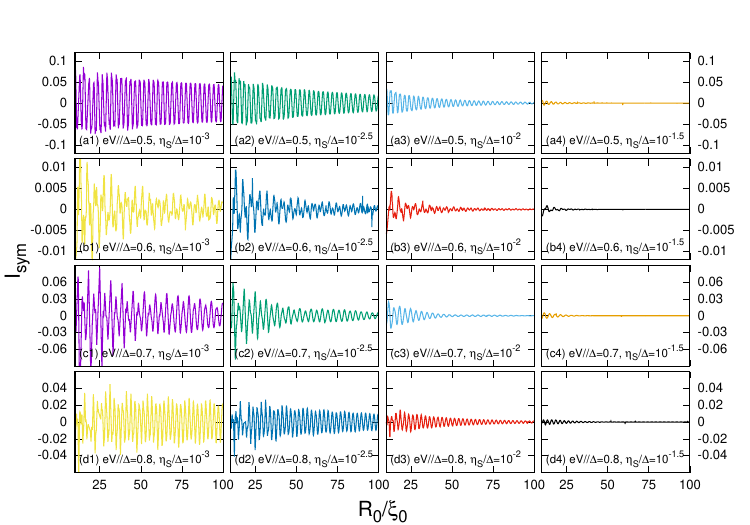}}
    \caption{{\it The effect of the Dynes parameter on the
        ultralong-distance effect:} The figure shows the symmetric
      current $I_{sym}$ defined by Eq.~(\ref{eq:Im}) for a
      $S_a$-$D_x$-$S_c$-$D_{x'}$-$S_b$ three-terminal Josephson
      junction biased at the voltages
      $(V_a,\,V_b,\,V_c)=(V,\,-V,\,0)$. Panels a1-a4, b1-b4, c1-c4 and
      d1-d4 correspond to
      $eV/\Delta=0.5,\,0.6,\,0.7,\,0.8$. respectively. The values of
      the Dynes parameters are the following: $\eta_S/\Delta=10^{-3}$
      (a1, b1, c1, d1), $\eta_S/\Delta=10^{-2.5}$ (a2, b2, c2, d2),
      $\eta_S/\Delta=10^{-2}$ (a3, b3, c3, d3) and
      $\eta_S/\Delta=10^{-1.5}$ (a4, b4, c4, d4).  The other
      parameters are identical to figure~\ref{fig:vs-R}. The current
      is in units of $e\Delta/\hbar$.
    \label{fig:vs-R-Dynes}
  }
\end{figure*}

\begin{figure*}[htb]
    \centerline{\includegraphics[width=.95\textwidth]{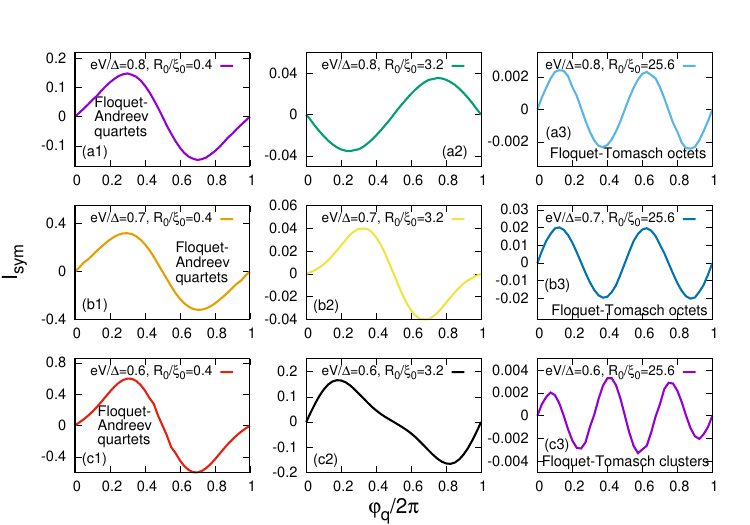}}
    \caption{{\it The cross-over from the Floquet-Andreev quartets to
        the ultralong-distance Floquet-Tomasch octets:} The figure
      shows the symmetric current $I_{sym}$ defined by
      Eq.~(\ref{eq:Im}) for a $S_a$-$D_x$-$S_c$-$D_{x'}$-$S_b$
      three-terminal Josephson junction biased at the voltages
      $(V_a,\,V_b,\,V_c)=(V,\,-V,\,0)$, as a function of
      $\varphi_q/2\pi$ on the $x$-axis. The voltage values are
      $eV/\Delta=0.5,\,0.4,\,0.3$ and
      $R_0/\xi_{ball}(0)=0.4,\,3.2,\,25.6$. The couplings between the
      quantum dots and the superconducting leads are the following:
      $\Gamma_{x,a}/\Delta = \Gamma_{x',b}/\Delta = 0.25$ and
      $\Gamma_{x,\alpha}/\Delta = \Gamma_{x',\beta}/\Delta = 1$. The
      Dynes parameter is $\eta_S / \Delta=10^{-3}${,
        thus with $l_\varphi^{(max)}=10^3 \xi_0$, where $\xi_0$ is a
        short notation for $\xi_{ball}(0)$.} The current is in units
      of $e\Delta/\hbar$.
    \label{fig:vs-V-1}
  }
\end{figure*}

\begin{figure*}[htb]
    \centerline{\includegraphics[width=.95\textwidth]{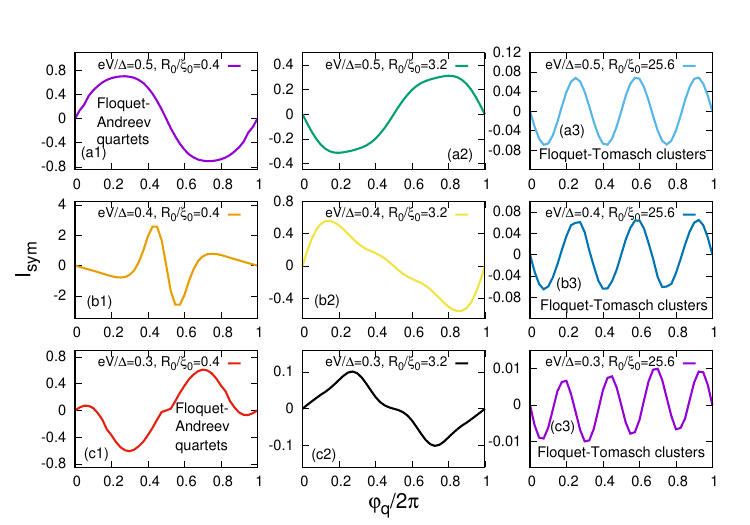}}
    \caption{{\it The cross-over from the Floquet-Andreev quartets to
        the ultralong-distance Floquet-Tomasch clusters:} The figure
      shows the symmetric current $I_{sym}$ defined by
      Eq.~(\ref{eq:Im}) for a $S_a$-$D_x$-$S_c$-$D_{x'}$-$S_b$
      three-terminal Josephson junction biased at the voltages
      $(V_a,\,V_b,\,V_c)=(V,\,-V,\,0)$, as a function of
      $\varphi_q/2\pi$ on the $x$-axis. The voltage values are
      $eV/\Delta=0.5,\,0.4,\,0.3$ and
      $R_0/\xi_{ball}(0)=0.4,\,3.2,\,25.6$. The couplings between the
      quantum dots and the superconducting leads are the following:
      $\Gamma_{x,a}/\Delta = \Gamma_{x',b}/\Delta = 0.25$ and
      $\Gamma_{x,\alpha}/\Delta = \Gamma_{x',\beta}/\Delta = 1$. The
      Dynes parameter is $\eta_S / \Delta=10^{-3}${,
        thus with $l_\varphi^{(max)}=10^3 \xi_0$, where $\xi_0$ is a
        short notation for $\xi_{ball}(0)$.}
    \label{fig:vs-V-2}
  }
\end{figure*}

Finally, we underline consistency with
Ref.~\onlinecite{McMillan-Anderson} regarding robustness with respect
to dephasing between the corresponding pairs of Green's function. The
superconducting Green's function $\hat{g}^A_{{\bf x},{\bf y}}$ in
Eq.~(\ref{eq:gA-supra-general-ballistique}) is rewritten as
\begin{eqnarray}
  \hat{g}^A_{{\bf x},{\bf y}}
  &=&
   \frac{1}{W}
    \frac{1}{k_F R} \exp\left\{\left(-\frac{R}{\xi_{ball}
      (\omega-i\eta_S)}\right)\right\}\times\\
    &&  \left[\cos\psi_F \hat{\cal M}_{cos}\left(\frac{\omega}{\Delta}\right)
    + \sin\psi_F \hat{\cal M}_{sin}\left(\frac{\omega}{\Delta}\right) \right]
      ,
\end{eqnarray}
where
\begin{eqnarray}
  \hat{\cal M}_{cos}\left(\frac{\omega}{\Delta}\right)
  &=&\frac{1}
  {\sqrt{\Delta^2-(\omega-i\eta_S)^2}}\times\\&&\nonumber
  \left(\begin{array}{cc} -(\omega-i \eta_S) & \Delta e^{i\varphi}\\
    \Delta e^{-i\varphi}&-(\omega-i\eta_S)\end{array}\right)\\
       \hat{\cal M}_{sin}\left(\frac{\omega}{\Delta}\right)
       &=&
       \left(\begin{array}{cc} -1 & 0 \\ 0 & 1
       \end{array}\right)
\end{eqnarray}
and $R=|{\bf x}-{\bf y}|$ is the distance between ${\bf x}$ and ${\bf
  y}$. We assume that the Fermi wave-length $\lambda_F=2\pi/k_F$ is
much smaller than all other length scales:
\begin{eqnarray}
\mbox{Re}\left[\frac{1}{\xi_{ball}
    (\omega-i\eta_S)}\right]\ll k_F\\
\mbox{Im}\left[\frac{1}{\xi_{ball}
    (\omega-i\eta_S)}\right]\ll k_F
.
\end{eqnarray}
In addition, the characteristic dimension $R_1$ of the quantum dot is
such that $R_1\ll \xi_{ball}(0)$, which implies that the
oscillations are not washed-out by extended contacts. Then, using the
notation
\begin{equation}
\langle\langle ... \rangle\rangle= \frac{k_F}{2\pi}
\int_{R_0}^{R_0+2\pi/k_F} dR
\end{equation}
for averaging over $R$ in the interval $\left[R_0,R_0+2\pi/k_F\right]$,
see Eq.~(\ref{eq:I-quartet-av}), we express the averaging
of the pairs of Nambu Green's functions as follows:
  \begin{widetext}
    \begin{eqnarray}
      \label{eq:averaging-modes}
    \langle\langle \hat{g}^A_{{\bf x},{\bf
        y}}\left(\frac{\omega_1}{\Delta}\right) \otimes
    \hat{g}^A_{{\bf y},{\bf
        x}}\left(\frac{\omega_2}{\Delta}\right)\rangle\rangle &\simeq&
    \frac{1}{2W^2} \frac{1}{(k_F R)^2}
    \exp\left\{\left(-\frac{R}{\xi_{ball}
      (\omega_1-i\eta_S)}\right)\right\}\exp\left\{\left(-\frac{R}{\xi_{ball}
      (\omega_2-i\eta_S)}\right)\right\}\times\\\nonumber &&\left[\hat{\cal
        M}_{cos}\left(\frac{\omega_1}{\Delta}\right)\otimes \hat{\cal
        M}_{cos}\left(\frac{\omega_2}{\Delta}\right) + \hat{\cal
        M}_{sin}\left(\frac{\omega_1}{\Delta}\right)\otimes \hat{\cal
        M}_{sin}\left(\frac{\omega_2}{\Delta}\right)\right] .
  \end{eqnarray}
  \end{widetext}
  The corresponding anomalous components involve one or two nonlocal
  Andreev electron-hole or hole-electron conversion. They take
  sizeable values if $\omega_1$, $\omega_2$ are typically in the
  energy window $0 < |\omega_1|,|\omega_2|\alt 2 \Delta$ instead of
  being strictly inside the gap according to $0 <
  |\omega_1|,|\omega_2|< \Delta$. This implies that the
  ultralong-distance effect holds for all of the quantum electron-hole
  conversion processes captured by Eq.~(\ref{eq:averaging-modes}), and
  being characterized by different sets of the corresponding 16 Nambu
  labels.  As a consequence, both the density of state oscillations of
  the two-terminal Tomasch effect and the clusters of Cooper pairs in
  the three-terminal Josephson junction are characterized by the
  corresponding ultralong-distance coupling, see also
  Appendix~\ref{app:connection} where the demonstration starts from
  the different point of view of the open boundary conditions
  considered by Wolfram and Lehman in Ref.~\onlinecite{Wolfram}.
  However, it is also shown in this Appendix~\ref{app:connection} that
  the coupling between the density of states at one contact and the
  pairs at the other contact is AC in the three-terminal Josephson
  junction. Thus, those AC density of state oscillations in a
  three-terminal Josephson junction cannot explain the following
  numerical data on the DC-current of clusters of Cooper pairs also
  with three superconducting terminals.
  
  To interpret the finite electron-hole or hole-electron conversion
  amplitude above the gap, in a characteristic spectral window
  $|\omega|\alt 2\Delta$, we refer to Fig. 7a in the
  Blonder-Tinkham-Klapwijk approach \cite{BTK}, showing the sizeable
  Andreev reflection conductance of a highly-transparent normal
  metal-superconductor junction as a function of voltage~$V$ such that
  $|eV|\alt 2\Delta$.

  In addition, the quasiparticles and pairs in the Tomasch
  oscillations in the two-terminal density of states
  \cite{Tomasch1,Tomasch2,Tomasch3} copropagate over
  ultralong-distance, and they can be referred to as ``the triplets
  correlations'' \cite{triplet} between a single quasiparticle and a
  pair. Conversely, two copropagating pairs correspond to ``the
  so-called quartets'' in three-terminal Josephson junctions
  \cite{Freyn}. A possibility is to speculate that enhanced
  condensation energy could be produced by those propagating Nambu
  modes acting like a ``glue'', in addition to the mean field BCS
  pairing. Indeed, it would be interesting to consider analogies with
  the theory of the collective modes
  \cite{collective1,collective2,collective3}, and to examine whether
  those ``triplets'' or ``quartets'' can possibly give rise to a
  collective state upon taking the Coulomb interaction or strong
  disorder into account.
  
\section{Numerical results}
\label{sec:numerical-results}

In this section, we provide selection of numerical data for the
reduced model~II, defined in the above section~\ref{sec:reduction-0D}. 

We successively introduce the calculations and present the
ultralong-distance effect, see figures~\ref{fig:vs-R}
and~\ref{fig:vs-R-Dynes}. {Next, we present} the cross-over from the
Floquet-Andreev quartets to the ultralong-distance Floquet-Tomasch
octets in the quartet phase sensitivity of the current, as the
distance between the dots is increased from $R_0/\xi_{ball}(0)\alt 1$
to $R_0/\xi_{ball}(0)\agt 1$ and to $R_0/\xi_{ball}(0)\gg 1$, see
figures~\ref{fig:vs-V-1} and~\ref{fig:vs-V-2}.

The current of the $S_a$-$D_x$-$S_c$-$D_{x'}$-$S_b$ double quantum dot
three-terminal Josephson junction is obtained from the fully dressed
Dyson-Keldysh equations to all orders in the tunneling
amplitudes. Concerning the algorithms, the code is based on
numerically exact implementation of the Dyson and Dyson-Keldysh
Eqs.~(\ref{eq:Dyson})-(\ref{eq:Gpm}), see
Appendix~\ref{app:themethod}. The Dyson Eq.~(\ref{eq:Dyson}) is solved
with recursive Green's functions in energy \cite{Cuevas} and sparse
matrix algorithms are used for the matrix products. Details about the
algorithms can be found in the Appendix of Ref.~\onlinecite{Sotto}.

Based on symmetry arguments \cite{Melin1,Jonckheere}, we implement the
hopping amplitudes $J_{a,x}=J_{b,x'}$ and $J_{\alpha,x}=J_{\beta,x'}$,
thus with $\Gamma_{x,a}=\Gamma_{x',b}$ and
$\Gamma_{x,\alpha}=\Gamma_{x',\beta}$ for the normal-state line-width
broadening parameters $\Gamma_n=J_n^2/W$. Then, we evaluate the current of
the clusters of Cooper pairs as
\begin{eqnarray}
  \label{eq:Im}
  I_{sym}\left(\frac{R_0}{\xi_{ball}(0)},\,\frac{eV}{\Delta},
  \frac{\varphi_q}{2\pi}\right)&=&I_a\left(\frac{R_0}{\xi_{ball}(0)},\,\frac{eV}{\Delta},
  \frac{\varphi_q}{2\pi}\right)\\ &+&I_b\left(\frac{R_0}{\xi_{ball}(0)},\,\frac{eV}{\Delta},
  \frac{\varphi_q}{2\pi}\right)
  ,
  \nonumber
\end{eqnarray}
where the currents $I_a$ and $I_b$ are transmitted into $S_a$ and
$S_b$, and $I_{sym}$ in Eq.~(\ref{eq:Im}) is averaged over $k_F R$
according to Eq.~(\ref{eq:I-quartet-av}).

We start with the sensitivity of $I_{sym}(R_0/\xi_{ball}(0),
eV/\Delta, \,\varphi_q/2\pi)$ on the distance $R_0/\xi_{ball}(0)$
between the quantum dots $D_x$ and $D_{x'}$. The data on
figure~\ref{fig:vs-R} show the current $I_{sym}$ as a function of
$R_0/\xi_{ball}(0)$ at the fixed quartet phase $\varphi_q/2\pi=0.1$
and for the reduced voltage values from $eV/\Delta=0.3$ to
$eV/\Delta=1.2$ on panels a-j respectively. The numerical data on
figure~\ref{fig:vs-R} feature complex pattern of the Floquet-Tomasch
oscillations. The beatings are interpreted as interference between the
wave-vectors of the quantum dot level Floquet replica. The numerical
data on figure~\ref{fig:vs-R} fully confirm the physical picture of
section~\ref{sec:physical-picture} regarding the ultralong-distance
Floquet-Tomasch oscillations. {The value $\eta_S/\Delta=10^{-3}$ of
  the Dynes parameter used in figure~\ref{fig:vs-R} implies
  $l_\varphi^{max}/\xi_{ball}(0)=10^3$, see
  Eqs.~(\ref{eq:l-phi-1})-(\ref{eq:l-phi-2}). This is compatible with
  emergence of sizeable $I_{sym}(R_0/\xi_{ball}(0), eV/\Delta,
  \,\varphi_q/2\pi)$ at $R_0/\xi_{ball}(0)=100$ on
  figure~\ref{fig:vs-R}. By contrast, $R_0$ is limited by $R_0\alt
  \xi_{ball}(0)$ in the recently considered Andreev molecules with all
  superconducting leads grounded
  \cite{Freyn,Pillet,Pillet2,Scherubl,Nazarov-PRR,Nazarov-PRB-AM}, and
  in the $F_aSF_b$ and $N_aSN_b$ Cooper pair beam splitters, see
  Refs.~~\onlinecite{exp-CPBS1,exp-CPBS2,exp-CPBS3,exp-CPBS4,exp-CPBS5,exp-CPBS6,exp-CPBS7,exp-CPBS8,theory-CPBS1,theory-CPBS2,theory-CPBS3,theory-CPBS4,theory-CPBS4-bis,theory-CPBS4-ter,theory-CPBS5,theory-CPBS6,theory-CPBS7,theory-CPBS8,theory-CPBS9,theory-CPBS10,theory-CPBS11,theory-noise8,theory-noise9,theory-noise11}.}
We also note that, strictly speaking, $\xi_{ball}(0)$ given by
Eq.~(\ref{eq:xi-ball-0}) and $l_\varphi^{max}$ given by
Eq.~(\ref{eq:l-phi-max}) are two independent length scales, in the
sense that $l_\varphi^{max}$ is not proportional to
$\xi_{ball}(0)$. The current $I_{sym}$ was averaged over the
oscillations at the scale of the Fermi wave-length
$\lambda_F=2\pi/k_F$ according to Eq.~(\ref{eq:I-quartet-av}). Then,
$\xi_{ball}(0)$ is the smallest length-scale to which the calculated
$I_{sym}$ couples and it is illustrative to plot $I_{sym}$ as a
function of $R_0/\xi_{ball}(0)$.

Figure~\ref{fig:vs-R-Dynes} illustrates the effect of the Dynes
parameter on the current $I_{sym}$. On this figure, the Dynes
parameter $\eta_S/\Delta$ ranges from $\eta_S/\Delta=10^{-3}$ (on
panels a1, b1, c1, d1) to $\eta_S/\Delta=10^{-2.5}$ (on panels a2, b2,
c2, d2), $\eta_S/\Delta=10^{-2}$ (on panels a3, b3, c3, 3) and
$\eta_S/\Delta=10^{-1.5}$ (on panels a4, b4, c4, d4). The voltage
values are $eV/\Delta=0.5,\,0.6,\,0.7,\,0.8$ on panels a1-a4, b1-b4,
c1-c4 and d1-d4 respectively.  It is concluded that the range of the
Floquet-Tomasch effect is strongly reduced by increasing the Dynes
parameter from $\eta_S/\Delta=10^{-3}$ to $\eta_S/\Delta=10^{-1.5}$,
in agreement with the physical arguments presented in the preceding
sections~\ref{sec:physical-picture}, \ref{sec:model-and-methods}
and~\ref{sec:reduction-0D}.

We also deduce from the $y$-scales on figure~\ref{fig:vs-R} that the
current $I_{sym}$ reaches maximum around $eV/\Delta\approx 1/2$, {\it
  i.e.} $I_{sym}$ for $eV/\Delta=0.4,\,0.5$ on figures~\ref{fig:vs-R}b
and~\ref{fig:vs-R}c is one order of magnitude larger than for
$eV/\Delta=0.3,\,0.6$ on figures~\ref{fig:vs-R}a
and~\ref{fig:vs-R}d. The strong enhancement of $I_{sym}$ at
$eV/\Delta=1/2$ is interpreted as the coinciding upper and lower gap
edge singularities of $S_a$ and $S_b$ which are biased at $\pm
V=\pm\Delta/2e$, as if the contact transparencies would be enhanced by
orders of magnitude in this voltage window{, see the remarks related
  to Eqs.~(\ref{eq:Iqc})-(\ref{eq:Ioc}) in the previous
  section~\ref{sec:reduction-0D}.}

It is also visible on figures~\ref{fig:vs-R-Dynes}
and~\ref{fig:vs-V-1} that the current $I_{sym}$ is larger for
$eV/\Delta=0.7$ than for $eV/\Delta=0.8$. The voltage-dependence of
$I_{sym}$ is indeed expected to be nonmonotonic, because of the
interplay between the voltage-$V$ sensitive peaks in the density of
states coming from the quantum dot Floquet replica, and the BCS gap
edge singularities, see the diagrams on figure~\ref{fig:octets}.

The cross-over from the Andreev quartets to the ultralong-distance
Floquet-Tomasch octets was proposed in
section~\ref{sec:physical-picture} as $R_0/\xi_{ball}(0)$ is increased
from $R_0/\xi_{ball}(0)\alt 1$ to $R_0/\xi_{ball}(0)\agt 1$ and next
to $R_0/\xi_{ball}(0)\gg 1$. Figures~\ref{fig:vs-V-1}
and~\ref{fig:vs-V-2} show how
$I_{sym}(R_0/\xi_{ball}(0),eV/\Delta,\,\varphi_q/2\pi)$ depends on the
quartet phase $\varphi_q/2\pi$ at fixed values of the reduced voltage
$eV/\Delta$ and distance $R_0/\xi_{ball}(0)$ between the quantum
dots. The values $eV/\Delta=0.5,\,0.4,\,0.3$ and
$R_0/\xi_{ball}(0)=0.4,\,3.2,\,25.6$ are used on
figure~\ref{fig:vs-V-1}, $eV/\Delta=0.5,\,0.4,\,0.3$ and
$R_0/\xi_{ball}(0)=0.4,\,3.2,\,25.6$ are used on
figure~\ref{fig:vs-V-2}.

In general, the symmetric current $I_{sym}$ has dominant quartet,
octet or higher-order $\varphi_q$-sensitivity, namely $I_{sym}\sim\sin
\varphi_q$, $I_{sym}\sim\sin (2\varphi_q)$ or $I_{sym}\sim\sin(n
\varphi_q)$ respectively.

The voltage $eV/\Delta=0.8$ on figures~\ref{fig:vs-V-1}-a1,
\ref{fig:vs-V-1}-a2 and {\ref{fig:vs-V-1}-a3 confirms}
the cross-over from the $\sin\varphi_q$ Andreev quartets to the
$\sin(2\varphi_q)$ ultralong-distance Floquet-Tomasch octets as
$R_0/\xi_{ball}(0)$ is increased from $R_0/\xi_{ball}(0)=0.8$ to
$R_0/\xi_{ball}(0)=3.2$ and to $R_0/\xi_{ball}(0)=25.6$. The dominant
$\sin \varphi_q$ and $-\sin\varphi_q$ are obtained for the small
$R_0/\xi_{ball}(0)=0.4$ and for the intermediate
$R_0/\xi_{ball}(0)=3.2$ on figures~\ref{fig:vs-V-1}-a1
and~\ref{fig:vs-V-1}-a2, while the dominant $\sin (2\varphi_q)$ of the
ultralong-distance Floquet-Tomasch octets is obtained for
$R_0/\xi_{ball}(0)=25.6$ on figure~\ref{fig:vs-V-1}-a3.

We also proposed in section~\ref{sec:physical-picture} emergence of
higher-order harmonics in the current-quartet phase relation as
$eV/\Delta$ is reduced. To illustrate this point, we now reduce the
bias voltage to $eV/\Delta=0.7$ (see figures~\ref{fig:vs-V-1}-b1,
\ref{fig:vs-V-1}-b2 and \ref{fig:vs-V-1}b3) and to $eV/\Delta=0.6$
(see figures~\ref{fig:vs-V-1}-c1, \ref{fig:vs-V-1}-c2 and
\ref{fig:vs-V-1}-c3). The following voltage values are also used on
figure~\ref{fig:vs-V-2}: $eV/\Delta=0.5$ (see
figures~\ref{fig:vs-V-2}-a1, \ref{fig:vs-V-2}-a2,
\ref{fig:vs-V-2}-a3), $eV/\Delta=0.4$ (see
figures~\ref{fig:vs-V-2}-b1, \ref{fig:vs-V-2}-b2, \ref{fig:vs-V-2}-b3)
and $eV/\Delta=0.3$ (see figures~\ref{fig:vs-V-2}-c1,
\ref{fig:vs-V-2}-c2, \ref{fig:vs-V-2}-c3). The dominant
$\sin(3\varphi_q)$ harmonics emerges for
$eV/\Delta=0.6,\,R_0/\xi_{ball}(0)=25.8$ on
figure~\ref{fig:vs-V-1}-c3, and for
$eV/\Delta=0.5,\,R_0/\xi_{ball}(0)=25.8$,
$eV/\Delta=0.4,\,R_0/\xi_{ball}(0)=25.8$ on
figures~\ref{fig:vs-V-2}-a3 and~\ref{fig:vs-V-2}-b3. The higher-order
$\sin(4\varphi_q)$ harmonics is also obtained for
$eV/\Delta=0.3,\,R_0/\xi_{ball}(0)=25.8$ on
figure~\ref{fig:vs-V-2}-c3.

We note consistency with our previous results for a single 0D quantum
dot \cite{Sotto}. Namely, $R_0/\xi_{ball}(0)=0.4$ and
$R_0/\xi_{ball}(0)=3.2$ on figures~\ref{fig:vs-V-1}-a1,
\ref{fig:vs-V-1}-a2 and~\ref{fig:vs-V-1}-c1 and on
figures~\ref{fig:vs-V-2}-a1, \ref{fig:vs-V-2}-a2
and~\ref{fig:vs-V-2}-c1 feature the $0$-to-$\pi$ and $\pi$-to-$0$
cross-overs which were found in our previous Ref.~\onlinecite{Sotto}

To summarize, the numerical calculations confirm the physical picture
of section~\ref{sec:physical-picture}, Appendix~\ref{app:connection}
and the analytical results of section~\ref{sec:reduction-0D} regarding
the following items: (i) The ultralong range of the effect and the way
it depends on the Dynes parameter ratio $\eta_S/\Delta$, (ii) The
sensitivity on the quartet phase $\varphi_q$, {\it i.e.} the
cross-over from the Andreev quartets to the ultralong-distance
Floquet-Tomasch octets as $R_0/\xi_{ball}(0)$ is increased from
$R_0/\xi_{ball}(0)\alt 1$ to $R_0/\xi_{ball}(0)\agt 1$ and to
$R_0/\xi_{ball}(0)\gg 1$, (iii) The voltage dependence of the effect,
{\it i.e.} the emergence of higher-order harmonics at smaller values
of the voltage $eV/\Delta$, and (iv) The emergence of large
ultralong-distance signal if $eV\simeq \pm\Delta/2$, which becomes
weaker if $eV$ is tuned away from $\pm \Delta/2$.

\section{Discussion}
\label{sec:perspectives}
In this section, {we discuss consequences for probing the
  ``quantumness'' of the Floquet-Tomasch clusters of Cooper pairs with
  quantum current-noise cross-correlations. We distinguish between
  theory (see subsection~\ref{sec:Sab-th}) and possible experiments
  (see section~\ref{sec:Sab-exp}).}

\subsection{Quantum current-noise cross-correlations}
\label{sec:Sab-th}
{The price to pay for nonlocal clusters of Cooper pairs over the
  ultralong-distance $R_0\sim l_\varphi$ is apparently to renounce to
  a ``good Floquet qu-bit''. Considering that the bias voltage energy
  $eV$ is much smaller than the superconducting gap $\Delta$, the
  Floquet resonance line-width broadening $\delta$ is limited by
  multiple Andreev reflections \cite{FWS,papierII,engineering,Berry},
  at least in the absence of ``extrinsic'' mechanism of relaxation
  \cite{FWS}.  We previously
  reported\cite{FWS,papierII,engineering,Berry} that $\delta \sim
  \exp(-c \Delta/eV)$ with $c$ of order unity, {\it i.e.}  the
  line-width broadening is exponentially small as $eV/\Delta$ is
  reduced. But here, the coupling to the continua of quasiparticles
  above the gap produces significant broadening of the Floquet
  resonances and small coherence time
  \cite{FWS,papierII,engineering,Berry} at higher voltage values, from
  $eV/\Delta=0.3$ to $eV/\Delta=1.2$ on figures~\ref{fig:vs-R},
  \ref{fig:vs-R-Dynes} and~\ref{fig:vs-V-1}.}

{This ``poor Floquet qu-bit'' does however not preclude
  emergence of quantum correlations at the ultralong distance $R_0\sim
  l_\varphi$, because the Cooper pair clusters are composite objects
  made of both the ``locally transmitted'' and ``nonlocally split''
  Cooper pairs, see figure~\ref{fig:device}d. It is known that, in
  general, breaking Cooper pairs produces quantum mechanical
  correlations and entanglement, see the $F_aSF_b$ and the $N_aSN_b$
  beam splitters
  \cite{theory-CPBS3,theory-CPBS4,theory-CPBS4-bis,theory-CPBS4-ter,theory-CPBS5,theory-CPBS6,theory-CPBS7,theory-CPBS8,theory-CPBS9,theory-CPBS10,theory-CPBS11,theory-noise8,theory-noise9,theory-noise11}. Nonvanishingly
  small zero-frequency quantum current-noise cross-correlations
  $S_{a,b}\ne 0$ in a $S_a$-dot-$S_c$-dot-$S_b$ three-terminal
  Josephson junction at the ultralong $R_0\sim l_\varphi$ is a
  possibility for experimental demonstration of the quantum nature of
  the ultralong-distance Cooper pair clusters.}

{In fact, the quantum current-noise cross-correlation
  kernel
\begin{equation}
S_{a,b}(\tau)=\hbar \int d\tau' K_{a,b}(\tau,\tau')
\end{equation}
was calculated by many authors, see for instance
Ref.~\onlinecite{Cuevas-noise} and Eqs.~(15)-(19) in our preceding
Ref.~\onlinecite{Sotto}:
\begin{eqnarray}
\nonumber
&&\hat{K}_{a,b}(\tau,\tau')=\frac{e^2}{\hbar^2}\mbox{Tr}\\
\label{eq:Sterme1}
&&\left\{\hat{J}_{\beta,b}(\tau) \hat{\tau}_3
\hat{G}^{+,-}_{b,a}(\tau,\tau')
\hat{J}_{a,\alpha}(\tau')\hat{\tau}_3
\hat{G}^{-,+}_{\alpha,\beta}(\tau',\tau)\right.\\
\label{eq:Sterme2}
&+&
\hat{J}_{b,\beta}(\tau) \hat{\tau}_3
\hat{G}^{+,-}_{\beta,\alpha}(\tau,\tau')
\hat{J}_{\alpha,a}(\tau') \hat{\tau}_3
\hat{G}^{-,+}_{a,b}(\tau',\tau)\\
\label{eq:Sterme3}
&-& \hat{J}_{\beta,b}(\tau)
\hat{\tau}_3 \hat{G}^{+,-}_{b,\alpha}(\tau,\tau')
\hat{J}_{\alpha,a}(\tau') \hat{\tau}_3
\hat{G}^{-,+}_{a,\beta}(\tau',\tau)\\
\label{eq:Sterme4}
&-& \hat{J}_{b,\beta}(\tau)\hat{\tau}_3
\hat{G}^{+,-}_{\beta,a}(\tau,\tau') \hat{J}_{a,\alpha}(\tau')
\hat{\tau}_3 \hat{G}^{-,+}_{\alpha,b}(\tau',\tau)\\ &&\left.+ (\tau
\leftrightarrow \tau')\right\}
\label{eq:ABC}
,
\end{eqnarray}
where $\hat{\tau}_3$ is a Pauli matrix, $\tau,\,\tau'$ are the time
variables and we assume $S_a$-$S_c$-$S_b$ three-terminal device which
is connected at the tight-binding sites $a$-$(\alpha,\beta)$-$b$ with
the hopping amplitudes $J_{a,\alpha} = J_{\alpha,a}$ and $J_{b,\beta}
= J_{\beta,b}$.  Eqs.~(\ref{eq:Sterme1})-(\ref{eq:ABC}) can be Fourier
transformed from the time variables $\tau,\,\tau'$ to the energies
$\omega+neV$ and $\omega+meV$, where $n$ and $m$ are two integers.}

{The nonvanishingly small current $I_{sym}\ne 0$ of the quartets,
  octets or higher-order clusters of Cooper pairs at the ultralong
  $R_0\sim l_\varphi$ implies nonvanishingly small Keldysh Green's
  functions $\hat{G}^{+,-}$ and $\hat{G}^{-,+}$, see the corresponding
  expressions of the current in Eq.~(\ref{eq:Im}),
  Eq.~(\ref{eq:I-a-alpha}) and
  Eqs.~(\ref{eq:I-spectral-1})-(\ref{eq:I-spectral-4}). Then,
  $S_{a,b}\ne 0$ at the ultralong $R_0\sim l_\varphi$ emerges on the
  condition that $\hat{G}^{+,-}$ and $\hat{G}^{-,+}$ in
  Eqs.~(\ref{eq:Sterme1})-(\ref{eq:ABC}) take values in
  overlapping energy intervals, {\it i.e.} the bias voltage $V\ne 0$
  should also be nonvanishingly small. In practice, the bias voltage
  energy $eV$ is a significant fraction of the superconducting gap
  $\Delta$.}

{Thus, within our model, the reported current
  $I_{sym}\ne 0$ implies quantum current-noise cross-correlations
  $S_{a,b}\ne 0$ due to the quantum fluctuations of the current operators
  at the ultralong-distance $R_0\sim l_\varphi$. Possible quantum
  noise cross-correlation experiments are considered now.}

\subsection{Proposed current cross-correlation experiments}
\label{sec:Sab-exp}

{On the experimental side, the positive zero-frequency quantum
  current-noise cross-correlations of the quartets were predicted
  \cite{Sotto} and measured in the Weizmann group experiment
  \cite{Heiblum}. In this experiment, absence of the quartet line and
  vanishingly small quantum current-noise cross-correlations
  $S_{a,b}=0$ were obtained with a pair of ``remote'' Josephson
  junctions. It was then concluded that ``the trivial effect'' of the
  electromagnetic environment is not at the origin of the quartet
  resonance line. The Grenoble experiment \cite{Lefloch} also ruled
  out ``extrinsic synchronization'' by demonstrating absence of the
  quartet line with remote contacts in a metallic structure.}

{The bias voltage was very low with respect to the superconducting gap
  in the Weizmann group experiment\cite{Heiblum}, {\it i.e.}
  $eV\ll\Delta$. Here, we propose analogous measurement of the quantum
  current-noise cross-correlations at voltage values that are
  significant fractions of the gap; typically $eV/\Delta$ is within
  the same range as on figures~\ref{fig:vs-R}, \ref{fig:vs-R-Dynes},
  \ref{fig:vs-V-1} and~\ref{fig:vs-V-2}, {\it i.e.}  from
  $eV/\Delta=0.3$ to $eV/\Delta=1.2$, given the above mentioned ``gap
  edge singularity resonance'' at $eV/\Delta=1/2$. We propose to
  systematically vary the distance $R_0$ between the junctions, in
  comparison with the superconducting coherence length $\xi_{ball}(0)$
  and the mesoscopic phase coherence length $l_\varphi$.  It is
  expected that the ultralong-distance Floquet-Tomasch clusters of
  Cooper pairs are above detection threshold, given the large signal
  in Tomasch experiment \cite{Tomasch1,Tomasch2,Tomasch3}.}

\section{Conclusions}
\label{sec:conclusions}

{Summary of the paper and final remarks are presented now.}

{We provided evidence} for ultralong-distance nonlocality and quantum
correlations in $S_a$-dot-$S_c$-dot-$S_b$ three-terminal Josephson
junctions where the constituting $S_a$-dot-$S_c$ and $S_c$-dot-$S_b$
are biased at opposite voltage on the quartet line. We presented
physical arguments in section~\ref{sec:physical-picture} and
Appendix~\ref{app:connection}, regarding the diagrammatic
interpretation of nonlocality. Analytical theory was proposed in
sections~\ref{sec:model-and-methods} and~\ref{sec:reduction-0D}. We
reduced the direct-gap semiconducting quantum dots to zero dimension,
and demonstrated emergence of the Floquet-Andreev and Floquet-Tomasch
currents limited by the relevant length scales of the superconducting
coherence length $\xi_{ball}(0)$ and the mesoscopic phase coherence
length of the superconducting quasiparticles $l_\varphi$
respectively. The numerical calculations presented in
section~\ref{sec:numerical-results} reveal that the ultralong-distance
Floquet-Tomasch clusters of Cooper pairs emerge if the separation
$R_0$ between the Josephson junctions exceeds the superconducting
coherence length~$\xi_{ball}(0)$ by orders of magnitude, {\it i.e.} if
$R_0\gg\xi_{ball}(0)$. This results from a phenomenological
description relying on the observation that the Dynes parameter
$\eta_S\ll \Delta$ is much smaller than the gap $\Delta$, which
implies that the corresponding mesoscopic phase coherence length
$l_\varphi\gg\xi_{ball}(0)$ of the superconducting quasiparticles is
much larger than the superconducting coherence length $\xi_{ball}(0)$.
In addition, in agreement with the physical arguments of
section~\ref{sec:physical-picture}, the voltage values are significant
fractions of the superconducting gap $\Delta$, typically
$eV>\Delta/2n$ for a cluster of order $n$, where $n$ is an
integer. The typical spectral window for the ultralong-distance effect
is roughly estimated as $|\omega|\approx 2\Delta$. Namely, the
ultralong-distance effect is obtained and nonlocal Andreev processes
are still sizeable if $|\omega|$ is not large compared to the
superconducting gap~$\Delta$. In this spectral window, the
superconducting quasiparticles behavior reflects both the normal- and
the superconducting-state properties. The numerical data confirm the
expectation that increasing $R_0/\xi_{ball}(0)$ from
$R_0/\xi_{ball}(0)\alt 1$ to $R_0/\xi_{ball}(0)\agt 1$ and to
$R_0/\xi_{ball}(0)\gg 1$ yields cross-over from the $\sin \varphi_q$
to the $\sin(2\varphi_q)$-sensitivities of the Floquet-Andreev
quartets and the ultralong-distance Floquet-Tomasch octets
respectively. Reducing $eV$ below $\Delta/2n$ produces
higher-order-$n$ clusters of Cooper pairs and dominant
$\sin(n \varphi_q)$ harmonics in the current, where $n$ is an integer.

The Tomasch oscillations \cite{Tomasch3} were experimentally observed
with superconducting film thickness $R_0$ as large as
$R_0=33.2\,\mu$m. Thus, in analogy with the Tomasch experiment
\cite{Tomasch3}, we conjecture emergence of the ultralong-distance
Floquet-Tomasch clusters of Cooper pairs if the separation between the
$S_a$-dot-$S_c$ and the $S_c$-dot-$S_b$ Josephson junctions is made as
large as $R_0=33.2\,\mu$m.

This predicted ultralong-range $R_0\sim l_\varphi$ of the
Floquet-Tomasch effect is spectacularly orders of magnitude above the
corresponding $R_0\sim\xi_{ball}(0)$ for overlapping Andreev bound
states at $V=0$
\cite{Pillet,Pillet2,Scherubl,Nazarov-PRR,Nazarov-PRB-AM} or for
$F_aSF_b$ or $N_aSN_b$ Cooper pair beam splitters, see
Refs.~\onlinecite{exp-CPBS1,exp-CPBS2,exp-CPBS3,exp-CPBS4,exp-CPBS5,exp-CPBS6,exp-CPBS7,exp-CPBS8,theory-CPBS1,theory-CPBS2,theory-CPBS3,theory-CPBS4,theory-CPBS4-bis,theory-CPBS4-ter,theory-CPBS5,theory-CPBS6,theory-CPBS7,theory-CPBS8,theory-CPBS9,theory-CPBS10,theory-CPBS11,theory-noise8,theory-noise9,theory-noise11}.

Finally, we show in Appendix~\ref{app:connection} that our numerical
experiments on the Floquet-Tomasch clusters of Cooper pairs and the
two-terminal density of state oscillations in the Tomasch experiments
\cite{Tomasch1,Tomasch2,Tomasch3} both involve ultralong-distance
behavior. But the microscopic processes are different, and, in a
three-terminal configuration, the coupling between the density of
states at one contact and the pairs at the other contact is AC and
thus, it cannot be proposed as an explanation for our numerical
experiments on the DC-current of the Cooper pair clusters.

To conclude, the length scale $l_\varphi$ for the mesoscopic phase
coherence of the superconducting quasiparticles was phenomenologically
introduced in our description. The effect offers the possibility to
directly probe quantum coherence of the superconducting quasiparticle
states, and to bridge with the physics of quasiparticle poisoning
\cite{Martinis2009,deVisser2011,Lenander2011,Rajauria2012,Wenner2013,Riste2013,LevensonFalk2014,Nazarov-qp},
in connection with the tremendous interest in the superconducting
circuits of quantum engineering. It seems that future experiments
could be a guideline towards further progress in understanding this
complex physics.  {Controlling the electromagnetic environment seems
  to be promising for producing small values of the Dynes parameter
  $\eta_S$ and long mesoscopic phase coherence $l_\varphi$ of the
  superconducting quasiparticles, see Ref.~\onlinecite{Pekola1}.}

\section*{Acknowledgements}

The author acknowledges the collaboration of the Weizmann Institute
group (Y.~Cohen, M.~Heiblum, Y.~Ronen, H.~Shtrikman) on interpretation
of the unpublished data which inspired this work. The author wishes to
thank B. Dou\c{c}ot for participating to the enjoying elaboration of
the framework of the interpretation.  The author also thanks
J.-G. Caputo and R. Danneau for discussions and critical reading of
the manuscript. The author wishes to thank \c{C}.\"{O}.~Girit,
J.D.~Pillet and their students and post-docs for sharing
Refs.~\onlinecite{Pillet,Pillet2} prior to making their preprint
publicly available. The author thanks the Centre R\'egional
Informatique et d'Applications Num\'eriques de Normandie (CRIANN) for
the use of its facilities. The author thanks the Infrastructure de
Calcul Intensif et de Donn\'ees (GRICAD) for the use of the resources of
the M\'esocentre de Calcul Intensif de l’Universit\'e Grenoble-Alpes
(CIMENT). The author acknowledges support from the French National
Research Agency (ANR) in the framework of the Graphmon project
(ANR-19-CE47-0007).

\appendix

\begin{figure*}[htb]
  \begin{minipage}{.7\textwidth}
    \includegraphics[width=.8\textwidth]{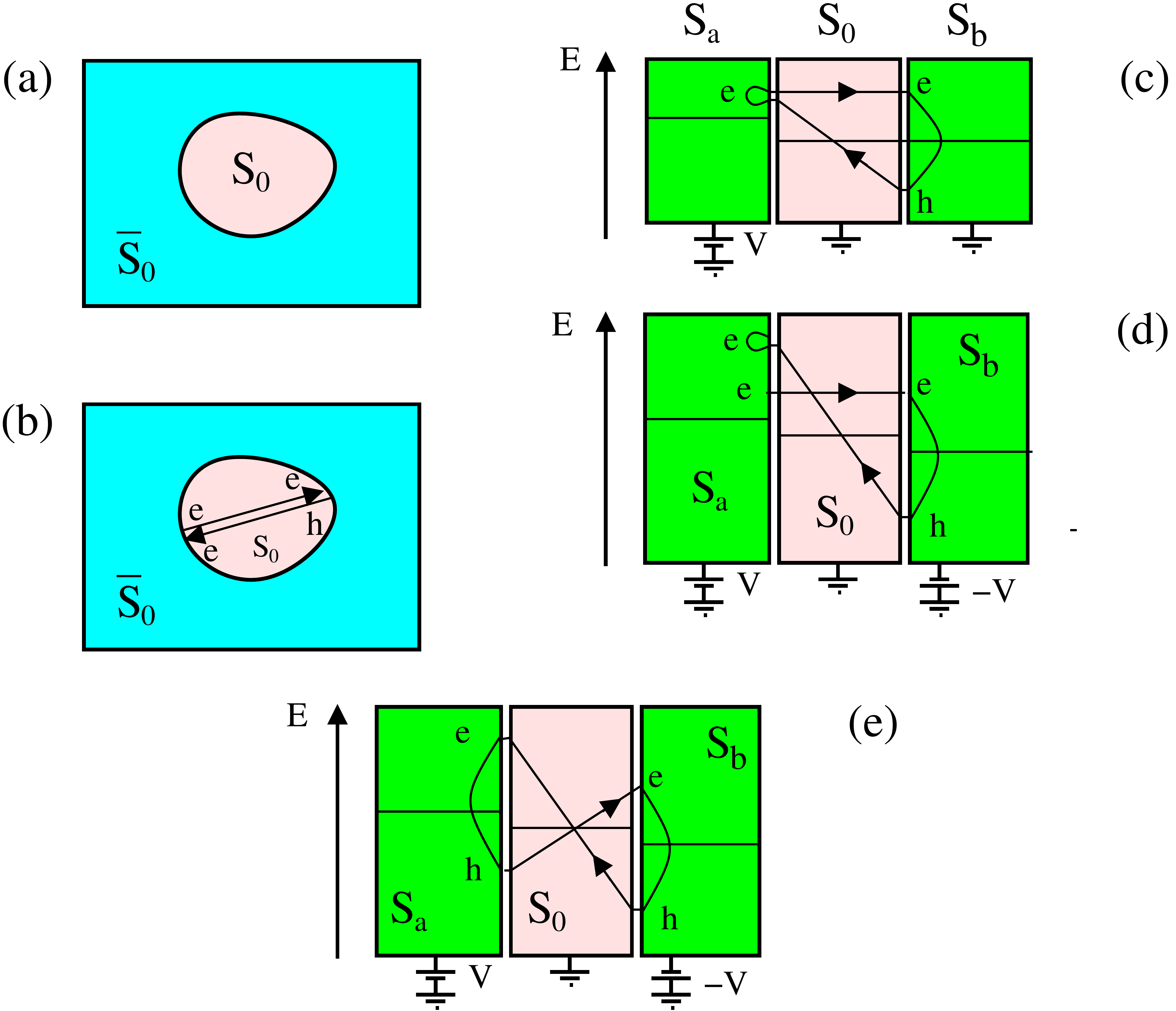}
    \end{minipage}\begin{minipage}{.29\textwidth}
  \caption{Panel a shows how a finite-size superconductor is defined
    in a bulk 3D superconductor, $S_0$ and $\overline{S}_0$ being the
    interior and the exterior respectively. Panel b shows nonlocality
    and a schematic representation of Eq.~(6) in
    Ref.~\onlinecite{McMillan-Anderson}. Panel c shows the
    ``triangular energy diagram'' for the DC-local density of states
    in a two-terminal configuration, as in the Tomasch experiment
    \cite{Tomasch1,Tomasch2,Tomasch3}. Panel d shows the AC-density of
    states in a three-terminal configuration, where $S_a$ and $S_b$
    are biased at opposite voltage while $S_0$ is grounded.  Panel e
    shows the three-terminal ``butterfly quartet energy diagram'' for
    the DC-transport of pairs and the Floquet-Tomasch effect, see also
    figure~\ref{fig:diagrams-current-and-noise}.
    \label{fig:schema-appendice}
  }\end{minipage}
\end{figure*}

\section{Connection with the Tomasch experiment}
\label{app:connection}
In this Appendix, we complement the main text by drawing a parallel
between the here considered nonlocal current-phase response of the
Floquet-Tomasch effect, and the density of state oscillations in the
Tomasch experiments \cite{Tomasch1,Tomasch2,Tomasch3}. We address this
question from two points of view: the ultralong-distance nonlocality
in section~\ref{sec:ULD} and the structure of the electron-hole
conversions in section~\ref{sec:Nambu}. This analogy further supports
the proposed interpretation of the numerical experiments in terms of
the diagrams that capture nonlocality, see
section~\ref{sec:physical-picture}. We justify in
section~\ref{sec:Nambu} the use of the vocabulary ``the
Floquet-Tomasch effect'' for the current of pairs in a three-terminal
Josephson junction. We also conclude to different quantum processes in
the density of state oscillations of the Tomasch effect
\cite{Tomasch1,Tomasch2,Tomasch3} and the current of pairs in a
three-terminal Josephson junction. Thus, the former cannot be used to
explain our calculations on the latter.

\subsection{Effects of a boundary}
\label{sec:ULD}
In this subsection, we start from a superconductor with open boundary
conditions, according to Wolfram and Lehman in
Ref.~\onlinecite{Wolfram}, and demonstrate that this implies
nonlocality in the sense of Eq.~(6) in
Ref.~\onlinecite{McMillan-Anderson} by McMillan and Anderson.

Namely, we consider that a finite-size region $S_0$ is defined in an
infinite 3D superconductor. The ``interior'' and the ``exterior'' are
denoted by $S_0$ an $\overline{S}_0$ respectively. Thus
$S_0+\overline{S}_0$ is an infinite 3D superconductor, see
figure~\ref{fig:schema-appendice}a.

We assume that the two-dimensional (2D) surface of $S_0$ is
practically realized with a collection of the Nambu hopping amplitudes
denoted by $\Sigma_{S_0,\overline{S}_0}$ and
$\Sigma_{\overline{S}_0,S_0}$ for hopping between $S_0$ and
$\overline{S}_0$ and between $\overline{S}_0$ and $S_0$
respectively. Those matrices $\Sigma_{S_0,\overline{S}_0}$ and
$\Sigma_{\overline{S}_0,S_0}$ have entries in the tight-binding sites
making the $S_0$-$\overline{S}_0$ interface and in the Nambu labels
({\it i.e.} they are diagonal in the Nambu labels).

We denote by $g$ and $G$ the Green's functions of $S_0+\overline{S}_0$
and $S_0$ respectively. We obtain $G$ for $S_0$ by including the
hopping self-energies
$\tilde{\Sigma}_{S_0,\overline{S}_0}=-\Sigma_{S_0,\overline{S}_0}$ and
$\tilde{\Sigma}_{\overline{S}_0,S_0}=-\Sigma_{\overline{S}_0,S_0}$
which cancel the plain 3D tight-binding amplitudes on the
$S_0$-$\overline{S}_0$ boundary. Thus, $S_0$ is disconnected from
$\overline{S}_0$ in the Green' function $G$ which is fully dressed
with the self-energy $\tilde{\Sigma}$.

The Dyson equations
  \begin{eqnarray}
&&    \left(I-g_{S_0,\overline{S}_0}
    \tilde{\Sigma}_{\overline{S}_0,S_0}\right) G_{S_0,S_0}
    -g_{S_0,S_0} \tilde{\Sigma}_{S_0,\overline{S}_0} G_{\overline{S}_0,S_0}
    = g_{S_0,S_0}\\
    &&
    -g_{\overline{S}_0,\overline{S}_0} \tilde{\Sigma}_{\overline{S}_0,S_0}
    G_{S_0,S_0}
    +
    \left(I-g_{\overline{S}_0,S_0}
    \tilde{\Sigma}_{S_0,\overline{S}_0}\right)
    G_{\overline{S}_0,S_0}
    =
    g_{\overline{S}_0,S_0}
    \end{eqnarray}
  have the following solution:
\begin{widetext}
  \begin{eqnarray}
    \label{eq:G-S0-S0}
      G_{S_0,S_0}&=&
      \left[I-g_{\overline{S}_0,\overline{S}_0}\tilde{\Sigma}_{\overline{S}_0,S_0}
        - g_{S_0,S_0} \tilde{\Sigma}_{S_0,\overline{S}_0}
        \left( I -g_{\overline{S}_0,S_0}\tilde{\Sigma}_{S_0,\overline{S}_0}\right)^{-1}
        g_{\overline{S}_0,\overline{S}_0}\tilde{\Sigma}_{\overline{S}_0,S_0}
        \right]^{-1} \times\\\nonumber&&
      \left[g_{S_0,S_0}
        +
        g_{S_0,S_0}\tilde{\Sigma}_{S_0,\overline{S}_0}
        \left( I -g_{\overline{S}_0,S_0}\tilde{\Sigma}_{S_0,\overline{S}_0}\right)^{-1}
        g_{\overline{S}_0,S_0}\right]
      .
    \end{eqnarray}
\end{widetext}

The density of states is sometimes called as ``the local density of
states'' because it can be measured with a local probe. It turns out
that the density of states in the Tomasch experiment nonlocally
couples to all of the thin-film boundary, if the conditions are met,
regarding the characteristic energy and length scales. Specifically,
we consider $R_0\alt l_\varphi$, where $R_0$ is the linear dimension
of $S_0$, see figures~\ref{fig:schema-appendice}a
and~\ref{fig:schema-appendice}b. In addition, we assume that the
energy is in the range $|\omega| \approx 2\Delta$, see the discussion
in section~\ref{sec:connection}. The phenomenological mesoscopic phase
coherence length $l_\varphi$ was introduced above in
section~\ref{sec:model-and-methods}. Then, Eq.~(\ref{eq:G-S0-S0})
implies that all pairs of tight-binding sites at the boundary of $S_0$
are connected to each other by the matrix $G_{S_0,S_0}$ taking roughly
similar order of magnitude for all pairs of sites at the boundary, on
the conditions $R_0\alt l_\varphi$ and $\omega\approx 2\Delta$.

Eq.~(\ref{eq:G-S0-S0}) also implies that conversion of spin-up
electron into spin-down hole (and vice-versa) is effective at the
boundary of $S_0$, which directly leads to Eq.~(6) in
Ref.~\onlinecite{McMillan-Anderson}, see also
figure~\ref{fig:schema-appendice}b. This implies compatibility of our
diagrammatic description with both
Refs.~\onlinecite{McMillan-Anderson,Wolfram}.

\subsection{The corresponding diagrams}
\label{sec:Nambu}
Now, we consider the electron-hole Nambu labels and examine a single
framework for deducing the different quantum processes that lead to
the Tomasch density of states oscillations
\cite{Tomasch1,Tomasch2,Tomasch3} and to the Floquet-Tomasch pair
current in three-terminal Josephson junctions. Those quantum processes
are characterized by the distinct diagrams on
figures~\ref{fig:schema-appendice}c, \ref{fig:schema-appendice}d and
Fig.~\ref{fig:schema-appendice}e.

Eq.~(6) in Ref.~\onlinecite{McMillan-Anderson} can schematically be
represented by the two-terminal ``triangular diagram'' on
figure~\ref{fig:schema-appendice}c. This quantum process involves
Andreev reflection at the thin-film boundary in the sense of spin-up
electron quasiparticle from $S_0$ being reflected as spin-down hole
quasiparticle in $S_0$. Then, a pair transmitted from the
quasiparticles states into the condensate of the same $S_0$, and the
crystal lattice has to be free to move in order to absorb the recoil
coming from conservation of momentum. The diagram on
figure~\ref{fig:schema-appendice}c involves electron-electron
propagation in the left superconductors $S_a$ and electron-hole
conversion in the right superconductor $S_b$. Thus,
figure~\ref{fig:schema-appendice}c encodes the Tomasch effect in the
sense Ref.~\onlinecite{McMillan-Anderson}, {\it i.e.}  the variations
of the density of states at the left interface as a function of the
electron-hole conversion at the other contact.

Conversely, figure~\ref{fig:schema-appendice}d shows schematically the
three-terminal diagram for the density of states. It does not form a
loop and thus, in a three-terminal configuration, the response in the
density of states at one contact in $S_a$ as a function of the pair
amplitude in $S_c$ features AC-oscillations.

Finally, the current of pairs in a double Josephson junction biased at
opposite voltages is captured by the ``quartet butterfly energy
diagram'' on figure~\ref{fig:schema-appendice}e, see also
Ref.~\onlinecite{Freyn} and
figures~\ref{fig:diagrams-current-and-noise}a
and~\ref{fig:diagrams-current-and-noise}b in
section~\ref{sec:physical-picture}. On
figure~\ref{fig:schema-appendice}e, two pairs are taken from $S_0$,
they exchange partners, a pair is transmitted into the left
superconductors $S_a$ in the final state, and another one into $S_b$
according to the quartet process \cite{Freyn}.

Thus, energy conservation implies that the ``triangular diagram
process'' on figure~\ref{fig:schema-appendice}c is DC in the
two-terminal configuration of the Tomasch experiment
\cite{Tomasch1,Tomasch2,Tomasch3}, but it becomes AC in the
three-terminal Josephson biased at opposite voltage. By contrast the
quartet diagram on figure~\ref{fig:schema-appendice}e is DC and this
is why our numerical calculations for the DC-ultralong-distance
Floquet-Tomasch current of pairs cannot be interpreted in terms of the
AC-density of state. Instead, they naturally receive the proposed
interpretation of the quartets and higher-order clusters of Cooper
pairs.

However, the straightforward wording of ``the Floquet-Tomasch effect''
is used throughout the paper for the three-terminal Josephson
junction, in order to refer to the common origin of the
ultralong-distance coupling in both cases.

\section{Details on the methods}
\label{app:themethod}

This subsection summarizes the method to evaluate the currents.

The calculation of the current \cite{Caroli,Cuevas} starts with
expression of the bare advanced and retarded Green's functions.

The bare Green's function of each quantum dot is given by
$g_{dot}(\omega)=\left(\omega-{\cal H}_{dot}-i\eta\right)^{-1}$, where
$\omega$ is the energy and ${\cal H}_{dot}$ is the quantum dot
Hamiltonian. Assuming the energy levels $\epsilon_\alpha$ and the
wave-functions $\langle {\bf x}|\alpha\rangle$ (at the location ${\bf
  x}$) yields the following electron-electron Green's function between
${\bf x}$ and ${\bf y}$:
\begin{eqnarray}
  \label{eq:g-rep-spec}
  g_{{\bf x},{\bf y}}^A(\omega)=\sum_\alpha \langle {\bf
    x}|\alpha\rangle \frac{1}{\omega-\epsilon_\alpha-\epsilon_g-i\eta}
  \langle \alpha|{\bf y} \rangle ,
\end{eqnarray}
where the gate voltage $V_g$-tunable energy $\epsilon_g$ fulfills the
condition $\epsilon_{\alpha_0}+\epsilon_g=0$ if $\alpha=\alpha_0$,
yielding resonance at zero energy $\omega=0$ (see
figure~\ref{fig:device} for the gates). Then, $g_{{\bf x},{\bf y}}^A(\omega)$ is
Eq.~(\ref{eq:g-rep-spec}) is approximated as
\begin{equation}
  g_{{\bf x},{\bf y}}^A(\omega)
  \simeq \frac{1}{\omega-i\eta}
  \langle {\bf
    x}|\alpha_0\rangle
\langle \alpha_0|{\bf y} \rangle
  \label{eq:g-rep-spec-approx}
.
\end{equation}
The parameter $\eta$ in Eq.~(\ref{eq:g-rep-spec}) is related to the
strength of relaxation.  It was found in Ref.~\onlinecite{FWS} that
tiny relaxation $0<\eta\ll\Delta$ has huge effect on the quartet
current, in comparison with the previous Ref.~\onlinecite{Jonckheere}
where $\eta=0^+$. However, the available experimental data
\cite{Heiblum} do not allow to demonstrate that $0<\eta\ll \Delta$ in
Ref.~\onlinecite{FWS} is more relevant than $\eta=0^+$ in
Ref.~\onlinecite{Jonckheere}. This is why the approximation $\eta=0^+$
is used in absence of further experimental input.

The $2\times 2$ Nambu representation has entries for spin-up electrons
and spin-down holes:
\begin{eqnarray}
    \label{eq:gA-Nambu}
    &&\hat{g}^A_{{\bf x},{\bf y}}(t,t')=-i\theta(t-t')\\
    \nonumber
   && \left(
    \begin{array}{cc}
    \langle \left\{c_{{\bf x},\,\uparrow}(t) , c^+_{{\bf
        y},\,\uparrow}(t') \right\}\rangle & \langle \left\{c_{{\bf
        x},\,\uparrow}(t) , c_{{\bf y},\,\downarrow}(t')
    \right\}\rangle\\ \langle \left\{c^+_{{\bf x},\,\downarrow}(t) ,
    c^+_{{\bf y},\,\uparrow}(t') \right\}\rangle & \langle
    \left\{c^+_{{\bf x},\,\downarrow}(t) , c_{{\bf y},\,\downarrow}(t')
    \right\}\rangle
    \end{array}
    \right) ,
  \end{eqnarray}
where $\langle \rangle$ denotes averaging in the stationary state,
$\{\}$ is an anticommutator, ${\bf x}$, ${\bf y}$ are the space
coordinates and $t$, $t'$ are the time variables.

Using Eq.~(\ref{eq:gA-Nambu}) and the Hamiltonian given by
Eqs.~(\ref{eq:H-BCS1})-(\ref{eq:H-BCS2}), we find the expression of
the bare superconducting Green's function with gap $\Delta$ and phase
$\varphi$:
\begin{eqnarray}
    \nonumber &&\hat{g}^A_{{\bf x},{\bf y}}(\omega)= \frac{1}{W}
    \frac{1}{k_F R} \exp\left\{\left(-\frac{R}{\xi_{ball}
      (\omega-i\eta_S)}\right)\right\}\\
\label{eq:gA-supra-general-ballistique}
&&  \left[\frac{\cos\psi_F}{\sqrt{\Delta^2-(\omega-i\eta_S)^2}}
  \left(\begin{array}{cc} -(\omega-i \eta_S) & \Delta e^{i\varphi}\\
    \Delta e^{-i\varphi}&-(\omega-i\eta_S)\end{array}\right)\right.\\
  &&\left. \left. + \sin\psi_F \left(\begin{array}{cc} -1 & 0 \\ 0 & 1
  \end{array}\right)\right]\right\}
  \nonumber
  ,
  \end{eqnarray}
where $R=|{\bf x}-{\bf y}|$ is the distance between ${\bf x}$ and
${\bf y}$ and $\varphi=\varphi_a,\,\varphi_b,\,\varphi_c$ according to
which of the $S_a,\,S_b$ or $S_c$ superconducting lead is
considered. The phase $\psi_F=k_F R$ in
Eq.~(\ref{eq:gA-supra-general-ballistique}) oscillates at the scale of
the small Fermi wave-length $\lambda_F=2\pi/k_F$, where $k_F$ is the
Fermi wave-vector. The ballistic {superconducting coherence} length
$\xi_{ball}$ at the energy $\omega$ is given by
Eq.~(\ref{eq:xi-ball-omega}).

Considering first vanishingly small
bias voltage $V=0$, the Nambu hopping amplitude connecting each
quantum dot to the superconductors takes the form
\begin{eqnarray}
  \hat{J}&=&\left(\begin{array}{cc} J_0 & 0 \\ 0 &
    -J_0 \end{array}\right)
  ,
\end{eqnarray}
where each contact has different $J_0$. For instance $J_0\equiv
J_{a,\alpha}$ at the $a$-$\alpha$ interface on
figure~\ref{fig:tight-binding}, and $J_0\equiv
J_{c,\gamma},\,J_{c',\gamma'}$ and $J_{b,\beta}$ at the $c$-$\gamma$,
$c'$-$\gamma'$ and $b$-$\beta$ interfaces.

The fully dressed advanced and retarded Nambu Green's functions
$\hat{G}^{A,R}$ are deduced from the bare ones by use of the Dyson
equation
\begin{equation}
  \label{eq:Dyson}
  \hat{G}^{A,R}=\hat{g}^{A,R}+\hat{g}^{A,R}
  \otimes \hat{J}\otimes \hat{G}^{A,R}
  ,
\end{equation}
where $\otimes$ denotes convolution over the time variables and
summation over the specific tight-binding sites at both ends of the
tunneling amplitude $\hat{J}$ connecting the dots to the
superconductors.

Assuming now voltage biasing on the quartet line according to
Eq.~(\ref{eq:quartet-line}), the superconducting phases
$\varphi_a(t)$, $\varphi_b(t)$ and $\varphi_c(t)$ of $S_a$, $S_b$ and
$S_c$ evolve according to the Josephson relations mentioned in the
Introduction. The overall quantum dynamics being time-periodic, the
Fourier-transformed Nambu Green's functions acquire the integer labels
$n,m$ regarding the harmonics $(2neV/\hbar,2meV/\hbar)$ of the
frequency $2eV/\hbar$ associated to the voltage $V$.

The fully dressed Keldysh Green's function $\hat{G}^{+,-}$ is given by
\cite{Caroli,Cuevas}
\begin{equation}
  \label{eq:Gpm}
  \hat{G}^{+,-}=\left(\hat{I}+\hat{G}^R \otimes \hat{J}
  \right) \otimes
  \hat{g}^{+,-}\otimes\left(\hat{I}+\hat{J} \otimes
  \hat{G}^A  \right) 
  ,
\end{equation}
where the bare Keldysh Green's function is
$\hat{g}^{+,-}(\omega)=n_F(\omega) \left[ \hat{g}^A(\omega) -
  \hat{g}^R(\omega)\right]$, with $n_F(\omega)$ the Fermi-Dirac
distribution function {\it i.e.}  $n_F(\omega)=\theta(-\omega)$ in the
limit of zero temperature, with $\theta(x)=1$ if $x>0$ and
$\theta(x)=0$ if $x<0$.

The current is next deduced from $\hat{G}^{+,-}$ given by
Eq.~(\ref{eq:Gpm}). For instance, the current through the $a-\alpha$
interface at time $t$ is given by \cite{Caroli,Cuevas}
\begin{eqnarray}
\label{eq:I-a-alpha}
  &&I_{a-\alpha}(t)=\\
  &&\frac{2e}{\hbar} \sum_p \left[ \hat{J}_{\alpha_p,a_p}
    \hat{G}^{+,-}_{a_p,\,\alpha_p}(t,t) - \hat{J}_{a_p,\,\alpha_p}
    \hat{G}^{+,-}_{\alpha_p,a_p}(t,t) \right]_{(1,1)}
\nonumber
  .
\end{eqnarray}
The subscript ``$(1,1)$'' in
Eq.~(\ref{eq:I-a-alpha}) stands for the electron-electron Nambu
component. Eq.~(\ref{eq:I-a-alpha}) can be expressed as
\begin{equation}
  I_{a-\alpha}= \frac{e}{\hbar} \int {\cal I}_{a,\alpha}(\omega) d\omega
  ,
\end{equation}
where the spectral current takes the form
\begin{eqnarray}
  \label{eq:I-spectral-1}
  {\cal I}_{a,\alpha}(\omega)&=& \sum_p \left[\left(
    \hat{J}_{\alpha_p,a_p}
    \hat{G}^{+,-}_{a_p,\,\alpha_p}\right)_{(1,1)/(0,0)}(\omega)\right.\\
&&    -\left( \hat{J}_{\alpha_p,a_p}
    \hat{G}^{+,-}_{a_p,\,\alpha_p}\right)_{(2,2)/(0,0)}(\omega)\\ &&-
    \left(\hat{J}_{a_p,\alpha_p}
    \hat{G}^{+,-}_{\alpha_p,\,a_p}\right)_{(1,1)/(0,0)}(\omega)\\
&&    +\left.\left(
    \hat{J}_{a_p,\alpha_p} \hat{G}^{+,-}_{\alpha_p,\,a_p}\right)_{(2,2)/(0,0)}(\omega)
    \right]
  .  \label{eq:I-spectral-4}
\end{eqnarray}
The subscripts ``(1,1)'' and ``(2,2)'' correspond to the
``electron-electron'' and ``hole-hole'' Nambu components and ``(0,0)''
encodes $n=m=0$ in the $(neV/\hbar, meV/\hbar)$ labels of the
harmonics of the Josephson frequency.

\section{Details on the analytical calculations}
\label{app:details}
Combining the Dyson Eq.~(\ref{eq:Dyson}) to
Eqs.~(\ref{eq:star-triangle1})-(\ref{eq:star-triangle2}) yields
\begin{eqnarray}
\label{eq:prop3}
G_{\alpha,\,\alpha}&=& \tilde{g}_{\alpha,x} \tilde{G}_{x,x} \tilde{g}_{x,\,\alpha}\\
\label{eq:prop4}
G_{\gamma,\,\alpha}&=& \tilde{g}_{\gamma,x} \tilde{G}_{x,x} \tilde{g}_{x,\,\alpha}
  .
\end{eqnarray}
The Dyson equations take the form
\begin{widetext}
  \begin{eqnarray}
    \label{eq:Dyson1}
    G_{\alpha,\,\alpha}&=& g_{\alpha,\,\alpha}+
    g_{\alpha,\,\alpha} \Sigma_{\alpha,a} g_{a,a} \Sigma_{a,\,\alpha} G_{\alpha,\,\alpha}
    +
    g_{\alpha,\,\gamma} \Sigma_{\gamma,c} g_{c,c} \Sigma_{c,\,\gamma} G_{\gamma,\,\alpha}
    +
    g_{\alpha,\,\gamma} \Sigma_{\gamma,c} g_{c,c'} \Sigma_{c',\,\gamma'} G_{\gamma',\,\alpha}\\
    \label{eq:Dyson2}
    G_{\gamma,\,\alpha}&=& g_{\gamma,\,\alpha}+
    g_{\gamma,\,\alpha} \Sigma_{\alpha,a} g_{a,a} \Sigma_{a,\,\alpha} G_{\alpha,\,\alpha}
    +
    g_{\gamma,\,\gamma} \Sigma_{\gamma,c} g_{c,c} \Sigma_{c,\,\gamma} G_{\gamma,\,\alpha}
    +
    g_{\gamma,\,\gamma} \Sigma_{\gamma,c} g_{c,c'} \Sigma_{c',\,\gamma'} G_{\gamma',\,\alpha}
    .
  \end{eqnarray}
\end{widetext}
Then, Eqs.~(\ref{eq:prop3})-(\ref{eq:prop4}) and Eq.~(\ref{eq:Dyson1})
yield
\begin{eqnarray}
  \label{eq:Dyson3}
  \tilde{G}_{x,x}&=&\tilde{g}_{x,x}+\tilde{g}_{x,x} \tilde{K}_{x,x}
  \tilde{G}_{x,x} + \tilde{g}_{x,x} \tilde{K}_{x,x'}
  \tilde{G}_{x',x}
  ,
\end{eqnarray}
where
\begin{eqnarray}
\label{eq:K1} \tilde{K}_{x,x}&=&
  \tilde{g}_{x,\,\alpha}\Sigma_{\alpha,a} g_{a,a} \Sigma_{a,\,\alpha}
  g_{\alpha,x}+ \tilde{g}_{x,\,\gamma}\Sigma_{\gamma,c} g_{c,c}
  \Sigma_{c,\,\gamma}
  \tilde{g}_{\gamma,x}\\\label{eq:K2}\tilde{K}_{x,x'}&=&
  \tilde{g}_{x,\,\gamma}\Sigma_{\gamma,c} g_{c,c'}
  \Sigma_{c',\,\gamma'} \tilde{g}_{\gamma',x'} .
\end{eqnarray}
Conversely, Eqs.~(\ref{eq:prop3})-(\ref{eq:prop4}) and
Eq.~(\ref{eq:Dyson2}) yield
\begin{eqnarray}
  \label{eq:Dyson4}
  \tilde{G}_{x,x}&=&\tilde{g}_{x,x}+\tilde{g}_{x,x} \tilde{K}'_{x,x}
  \tilde{G}_{x,x} + \tilde{g}_{x,x}+\tilde{g}_{x,x} \tilde{K}'_{x,x'}
  \tilde{G}_{x',x}
  ,
\end{eqnarray}
where it turns out that $\tilde{K}'_{x,x}=\tilde{K}_{x,x}$ and
$\tilde{K}'_{x,x'}=\tilde{K}_{x,x'}$. Thus, Eqs.~(\ref{eq:Dyson3})
and~(\ref{eq:Dyson4}) are compatible with each other.  Given
Eq.~(\ref{eq:Gpm}),
Eqs.~(\ref{eq:I-spectral-1})-(\ref{eq:I-spectral-4}) and
Eq.~(\ref{eq:prop3}), we obtain
\begin{eqnarray}
  && \Sigma_{a,\,\alpha} G^{+,-}_{\alpha,a}\\ &=&
  \left(\Sigma_{a,\,\alpha} \tilde{g}_{\alpha,x} \tilde{G}_{x,x}
  \tilde{g}_{x,x} \Sigma_{\alpha,a} g_{a,a}\right)^{+,-}\\ &=&
  \Sigma_{a,\,\alpha} \tilde{g}^{+,-}_{\alpha,x} \tilde{G}^A_{x,x}
  \tilde{g}^A_{x,x} \Sigma_{\alpha,a} g^A_{a,a} \\&+&
  \Sigma_{a,\,\alpha} \tilde{g}^{R}_{\alpha,x} \tilde{G}^{+,-}_{x,x}
  \tilde{g}^A_{x,x} \Sigma_{\alpha,a} g^A_{a,a}\\ &+&
  \Sigma_{a,\,\alpha} \tilde{g}^{R}_{\alpha,x} \tilde{G}^{R}_{x,x}
  \tilde{g}^{+,-}_{x,x} \Sigma_{\alpha,a} g^A_{a,a}\\ &+&
  \Sigma_{a,\,\alpha} \tilde{g}^{R}_{\alpha,x} \tilde{G}^{R}_{x,x}
  \tilde{g}^{R}_{x,x} \Sigma_{\alpha,a} g^{+,-}_{a,a} .
\end{eqnarray}
The Dyson Eqs.~(\ref{eq:Dyson1})-(\ref{eq:Dyson2})
\begin{eqnarray}
  \label{eq:Dy1}
  \tilde{G}_{x,x}&=& \tilde{g}_{x,x}+
  \tilde{g}_{x,x} \tilde{K}_{x,x} \tilde{G}_{x,x} +
  \tilde{g}_{x,x} \tilde{K}_{x,x'} \tilde{G}_{x',x}\\
  \tilde{G}_{x',x}&=& 
  \tilde{g}_{x',x'} \tilde{K}_{x',x} \tilde{G}_{x,x} +
  \tilde{g}_{x',x'} \tilde{K}_{x',x'} \tilde{G}_{x',x}
  \label{eq:Dy2}
\end{eqnarray}
lead to
\begin{equation}
  \label{eq:Gxx}
  \tilde{G}_{x,x}=\left[I- \tilde{L}_{x,x} \tilde{K}_{x,x'}
    \tilde{L}_{x',x'} \tilde{K}_{x',x} \right]^{-1} \tilde{L}_{x,x} ,
\end{equation}
where
\begin{eqnarray}
  \tilde{L}_{x,x}&=& \left[ \left(\tilde{g}_{x,x}\right)^{-1}-\tilde{K}_{x,x}\right]^{-1}\\
  \tilde{L}_{x',x'}&=& \left[ \left(\tilde{g}_{x',x'}\right)^{-1}-\tilde{K}_{x',x'}\right]^{-1}
  .
\end{eqnarray}
Then, we deduce Eqs.~(\ref{eq:Dyson-exp})-(\ref{eq:Dyson-exp4}) in
section~\ref{sec:reduction-0D}.

\end{document}